\documentclass[journal]{IEEEtran}

\usepackage{booktabs}
\usepackage{amsfonts}
\usepackage{placeins}
\usepackage{acronym}
\usepackage{amsmath}
\usepackage{tabularx}
\usepackage{stfloats}
\usepackage{siunitx}
\normalsize
\usepackage{amsmath,amssymb,amsfonts}
\usepackage{subcaption}
\usepackage{setspace} 
\usepackage{mathtools}
\usepackage{color}
\usepackage{array}
\usepackage[version=4]{mhchem}
\usepackage{float}
\usepackage{stmaryrd}
\usepackage{soul}
\usepackage{stackrel}
\usepackage{epstopdf}
\usepackage{url}
\usepackage{cite}
\usepackage{tikz}
\usetikzlibrary{automata,arrows,positioning,calc}
\usepackage{bigints}
\usepackage{booktabs}
\usepackage{threeparttable}
\usepackage{bbm}
\usepackage{dsfont}
\usepackage{pgfplots}\pgfplotsset{compat=1.18}
\usepackage{tikz}
\usetikzlibrary{arrows.meta, positioning}
\usetikzlibrary{angles,quotes} 
\usetikzlibrary{shadings}
\newcolumntype{L}[1]{>{\raggedright\arraybackslash}m{#1}}
\newcolumntype{C}[1]{>{\centering\arraybackslash}m{#1}}
\acrodef{1D}{one-dimensional}
\acrodef{2D}{two-dimensional}
\acrodef{3D}{three-dimensional}
\acrodef{ODE}{ordinary differential equation}
\acrodef{PDE}{partial differential equation}
\acrodef{TX}{transmitter}
\acrodef{RX}{receiver}
\acrodef{MC}{molecular communication}
\acrodef{LEGI}{local excitation global inhibition}
\acrodef{BAA}{Blahut-Arimoto algorithm}
\acrodef{IBAA}{inverse Blahut-Arimoto algorithm}
\acrodef{cryo-EM}{cryo-electron microscopy}
\acrodef{MCP}{methyl-accepting chemotaxis protein}
\acrodef{GPCR}{G protein coupled receptor}
\acrodef{PI3K}{phosphoinositide 3-kinase}
\acrodef{PIP3}{phosphatidylinositol (3,4,5)-trisphosphate}
\acrodef{GTPase}{guanosine griphosphate hydrolase}
\acrodef{FRET}{Förster resonance energy transfer}
\acrodef{TIRF}{total internal reflection fluorescence}
\acrodef{cAMP}{cyclic adenosine monophosphat}
\acrodef{RDT}{rate distortion theory}
\acrodef{PMF}{probability mass function}
\acrodef{IRDT}{indirect rate distortion theory}
\acrodef{c8}[C$8^{\star}$]{caspase 8}

\ifCLASSINFOpdf
\else
\fi

\begin{document}

\title{Inferring the Chemotaxis Distortion Function from Cellular Decision Strategies}
\author{
    Fardad Vakilipoor,
    Johannes Konrad,
    Maximilian Sch\"afer
\thanks{F. Vakilipoor, J. Konrad, and M. Sch\"afer are with the Institute for Digital Communications, Friedrich-Alexander-Universität Erlangen-Nürnberg, Germany (fardad.vakilipoor@fau.de).}
}
%
%

\markboth{Inferring the Chemotaxis Distortion Function from Cellular Decision Strategies}%
{Konrad \MakeLowercase{\emph{et al.}}}

\maketitle

\newcommand{\rev}[1]{\textcolor{red}{#1}}

\begin{abstract}
Cellular intelligence enables cells to process environmental signals and make context-dependent decisions, as exemplified by chemotaxis, where cells navigate chemical gradients despite noisy signaling pathways. To investigate how cells deal with uncertainty, we apply an information-theoretic framework based on \ac{RDT}. The \ac{BAA} computes optimal decision strategies that minimize mutual information while satisfying distortion constraints, balancing sensing accuracy with distortion constraint equivalent to resource cost. We propose the \ac{IBAA} to compute the distortion function, which quantifies the system’s decision-making criteria for realizing a decision strategy to map input signals to outputs. This general framework extends beyond chemotaxis to biological and engineered systems requiring efficient information processing under uncertainty. We validate the proposed \ac{IBAA} by accurately estimating theoretical distortion functions in a cellular apoptosis scenario. Additionally, using the \ac{LEGI} model to simulate chemotactic responses, we compute the distortion functions from the cell's perspective. Our finding reveals a state-dependent decision criteria by the cell.
\end{abstract}
\IEEEpeerreviewmaketitle
\section{Introduction}
Communication is a fundamental feature of nature that enables interactions across all levels of life and matter. From subatomic particles exchanging energy to organisms transmitting signals vital for survival, information propagates to regulate and align behavior. In nature, communication takes various forms: plants release volatiles to warn neighboring plants of herbivore attacks, birds and insects signal territory or mating readiness through song and color displays, and animals rely on electrical, chemical, or mechanical signals to convey essential messages~\cite{vakilipoor2025agri,Echeverri2021,Kota2019,Kaplan2014}. At smaller scales, cells exchange signals to coordinate development, maintain homeostasis\footnote{Homeostasis describes the ability of an organism or a cell to maintain stable internal conditions, even when the external environment changes.}, and respond/adapt to environmental changes~\cite{Bi2021}. Across scales, from molecules to organisms, communication is a unifying process that ensures the proper functioning of biological systems.

At the core of these processes is the concept of information, defined as the reduction of uncertainty when a signal is received~\cite{Tarek2024}. Cells act as information-processing systems, employing interconnected biochemical pathways to detect cues, transduce signals, and generate responses. Neurons integrate synaptic inputs into precisely timed electrical signals for rapid nervous-system communication~\cite{Kandel2000}; immune cells detect rare pathogen-derived molecules amidst noise with remarkable precision~\cite{Janeway1999}; and embryonic cells interpret morphogen gradients to establish spatial patterns during development~\cite{Wolpert2002}. In each case, cells balance detection accuracy and decision-making against noise, energy, and physical limits. Understanding how cells gather, process, and act upon information remains a central challenge at the interface of biology, physics, and engineering, with implications for neuroscience, immunology, synthetic biology, and bioengineering~\cite{Fairhall2001,Thierry2010,Daniel2013,Asthagiri2000}.

\begin{figure*}[t]
    \centering
    \includegraphics[width=0.9\linewidth]{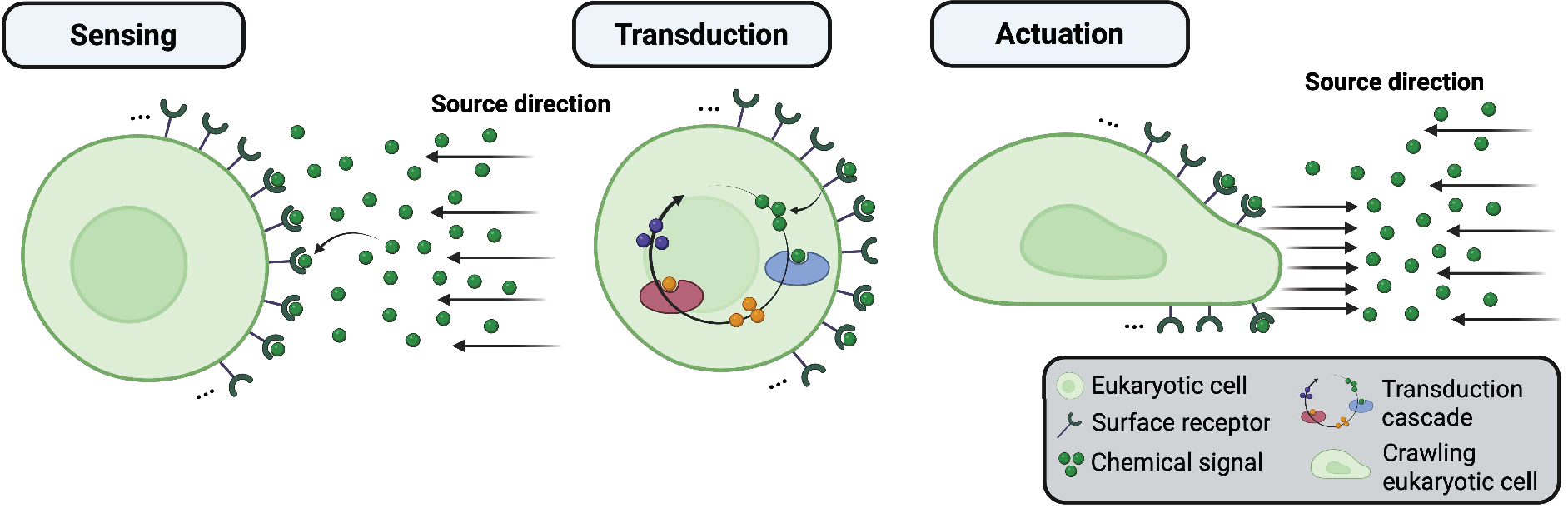}
    \caption{\small Modular view of the chemotaxis process three stages. From left to right: sensing, signal transduction, and actuation stage (created with biorender.com).}
    \label{fig:schematic}
\end{figure*}
One example of cellular information processing is chemotaxis. Chemotaxis is defined as the directed movement of cells in response to gradients of chemical substances known as chemoattractants~\cite{Monica2014}. Many cell types, including bacteria, immune cells, and single-celled eukaryotes, perform chemotaxis for critical tasks such as locating nutrients, navigating toward infected sites, or participating in tissue development~\cite{Pan2016,Paliwal2007,JUKAM20106}. 
The complete process of chemotaxis consists of a number of separable, but interrelated stages. The first is the \textit{sensing stage}, whereby an environmental signal is detected by the cells. In the literature, this stage is referred to by various terms including signal detection, ligand binding, stimulus perception, receptor activation, or environmental cue recognition~\cite{Miller2004}. The second stage is the \textit{decision stage}, which is also known as signal transduction, adaptation mechanisms, gradient interpretation, or sensory integration. The third and final stage is the \textit{actuation stage} where a response to the environmental signal is produced. This stage is known as motility regulation, motor response, directional movement, cytoskeletal reorganization, or effector activation (see Fig.~\ref{fig:schematic}). As each of the individual stages involve molecular interactions (e.g., receptor bindings, chemical reactions, etc.), the transmitted, processed, and received signals are subject to stochastic fluctuations~\cite{Ueda2007}. Stochastic fluctuations are an important fact of life in biological systems and they have been investigated from the earliest days of systems biology~\cite{McAdams_1997}. For example, cell migration in the amoeba Dictyostelium discoideum is assumed to be largely driven by stochastic triggering of an excitable system~\cite{Matsuoka2024}. 

Extensive research has been dedicated to developing theoretical frameworks that characterize noisy cellular signaling~\cite{Raymond2011}, along with simulation techniques that capture intrinsic stochastic variability~\cite{Gillespie_2007}, and experimental methodologies that elucidate the functional consequences of such noise~\cite{Michael_2002}. In order to gain deeper insights into the functioning of biological systems under noisy conditions, also engineering-inspired approaches have been increasingly employed. For instance, several studies have investigated how cells implement intrinsic noise-filtering mechanisms~\cite{Andrews_2006,Chou2011,Zechner_2016}. Synthetic biology has enabled the engineering of molecular filters that operate within living cells~\cite{Laurenti2018}. Noisy cellular signaling can also be investigated through the lens of information theory, particularly within the framework of Shannon’s communication theory. Over the past decade, numerous reviews have examined this perspective~\cite{Perkins_2009, Mian_2011, Tkacik_2011, Rhee_2012, Wolde2025, MOUSAVIAN20163, MOUSAVIAN201614, UDA201624}.
While much of these works emphasize channel capacity, in this work we focus on \acf{RDT}, which has emerged as a more biologically grounded method for analyzing how signaling networks function as information processing systems~\cite{Andrews2007,Porter2012, Iglesias2016}. \Ac{RDT} provides bounds on the rate at which information must be transmitted through a system to meet a given performance criterion. Authors in~\cite{Andrews2007} reformulated the response of a cell as an optimal rate distortion problem, examining the amount of information about the external signal that a cell requires to migrate directionally with a specified level of accuracy. In~\cite{Porter2012}, authors focused on binary decision making by cells based on the noisy signals received from their environment. They investigated cellular decision-making strategies and showed that several observed behaviors in binary systems, including random strategies, hysteresis, and irreversibility, are optimal from an information-theoretic perspective. In~\cite{Iglesias2016}, authors extended the idea of using \ac{RDT} and decision-making strategies to a broader group of systems. They showed how several stimulus response curves that are frequently observed in biological signaling pathways arise naturally as the optimal decision strategy based on \ac{RDT}. 

Extending previous works, this work investigates how information theory can shed light on the criteria that cells use to navigate in noisy (chemical) environments\footnote{In this context, noise refers to the stochastic fluctuations and uncertainties in the sensed chemical signals, which can arise from molecular diffusion, receptor binding randomness, or dynamic environmental changes.}. In this work we assume that cellular mechanisms in nature have evolved toward optimal decision strategies. From \ac{RDT} it is understood that it is possible to maintain an optimal decision strategy based on a certain criterion, namely the distortion function. In this work, we propose the \acf{IBAA} and reformulate the \ac{RDT} problem by identifying, from the system's perspective, the distortion function that yields a specific decision strategy. We validate the proposed \ac{IBAA} using an apoptosis model and assess its ability to estimate the theoretically defined distortion functions. Furthermore, by modeling cellular sensing with the established \acf{LEGI} framework, we apply the \ac{IBAA} to identify the distortion function that reflects the cell’s internal criterion for extracting directional information from chemical gradients. Hence, we attempt to understand the principle underlying the cell's decision strategy. This approach not only clarifies how cells optimize chemotactic responses under uncertainty, but also provides a framework for inferring implicit distortion functions in other biological systems, such as collective navigation in birds or foraging in bees. More broadly, these insights contribute to uncovering general principles that govern decision-making and information processing across living systems.

The main contributions of this paper are as follows:
\begin{itemize}
    \item We derive the transition probability distribution for the chemotaxis process using the \ac{LEGI} model. The derived transition probability is interpreted as the decision strategy of an information processing system.
    \item We employ \ac{RDT} to compute optimal decision strategies that minimize mutual information while satisfying distortion constraints, using the \acf{BAA}.
    \item We propose \ac{IBAA} to compute the distortion function, which results in the system’s decision strategy for mapping inputs to outputs.
    \item We validate the \ac{IBAA} in a binary decision process, i.e., in cellular apoptosis, demonstrating its capability to accurately reconstruct the theoretical distortion functions considered in the simulation process.
    \item We estimate the distortion function for the chemotaxis process according to the \ac{LEGI} model with the proposed \ac{IBAA}, and demonstrate that higher distortion values (i.e., stronger penalization) correspond to higher Hill coefficients, indicating that the cell’s internal state shapes the distortion function’s when greater accuracy is required.
\end{itemize}

All code and data supporting the simulations and evaluations in this work are publicly available in the repository at~\cite{zenodo_17193165}.
The remainder of this paper is organized as follows:
In Section~\ref{sec:chemotaxis}, we introduce the fundamental concepts of chemotaxis, provide representative examples from nature, and summarize key research directions pertaining to chemotaxis process.
In Section~\ref{sec:system_model}, we model the chemotaxis process based on the \ac{LEGI} mechanism. 
In Section~\ref{sec:RD_Theory}, we introduce the concept of \ac{RDT} necessary to analyze information processing and explain \ac{BAA} to obtain the optimum decision strategy knowing the theoretical distortion function. In Section~\ref{sec:IBAA}, we propose \ac{IBAA}, which is used to estimate the distortion function based on the input output data.
Section~\ref{sec:results} presents the results, including the investigation on two cases of apoptosis and chemotaxis. Finally, Section~\ref{sec:conclusion} concludes the paper and discusses potential directions for future work.

\section{Introduction to Chemotaxis}\label{sec:chemotaxis}
 In this section, we introduce chemotaxis in biological systems by reviewing various examples and key research directions. Chemotaxis is a fundamental biological process conserved across diverse organisms and described the directed movement of cells in response to chemical gradients~\cite{VanHaastert2004}. It enables key biological functions such as immune cell migration to infection sites, fibroblast recruitment during wound healing, and nutrient-seeking behaviors in unicellular organisms like bacteria and amoebae~\cite{Stock2009}. Table~\ref{tab:chemotaxis_examples} summarizes various examples for chemotaxis in biological systems. Beyond its biological significance, chemotaxis also has broad applications: in medicine, understanding chemotactic signaling guides targeted drug delivery to diseased tissues~\cite{Li2020}; in synthetic biology, engineered cells with tailored responses act as living therapeutics that seek out pathogens or deliver localized treatments~\cite{Riglar2018}; and in biotechnology, chemotaxis-inspired principles support microrobot design for navigating within complex environments by following chemical gradients, with potential applications in environmental sensing and minimally invasive surgery~\cite{Zarepour2024}.

Chemotaxis research spans various domains: molecular and cellular studies explore signaling networks and regulatory mechanisms controlling chemotactic behavior~\cite{Snyderman1981}; structural and biophysical studies use techniques such as \ac{cryo-EM} and microfluidics to investigate spatial organization and mechanical forces in signal propagation~\cite{Briegel2012}; mathematical and computational models simulate gradient sensing and motility across scales~\cite{Herrero2007}; information-theoretic frameworks analyze how noise and energy constraints impact the accuracy and efficiency of gradient sensing and decision-making~\cite{Mattingly2021}.

\begin{table*}[!ht]
\scriptsize
\caption{\small Examples of chemotaxis across biological systems.}
\begin{tabular}{|
    >{\centering\arraybackslash}m{2.3cm} |
    >{\centering\arraybackslash}m{2.5cm} |
    >{\centering\arraybackslash}m{6cm} |
    p{5.55cm} |}
\hline
\textbf{Cell Type} & \textbf{Environment} & \textbf{Signal Types (simplified)} & \multicolumn{1}{c|}{\textbf{Mechanism}} \\
\hline
Escherichia coli & Aquatic or host environments & Nutrients (e.g., amino acids, sugars) & Movement by alternating runs and tumbles, controlled by flagella in response to detected chemicals~\cite{Wadhams2004}. \\
\hline
Neutrophils & Sites of infection or inflammation & Immune signals released by bacteria or damaged tissue & Directed migration toward infection sites to eliminate pathogens~\cite{Kolaczkowska2013}. \\
\hline
Dictyostelium discoideum & Soil with nutrient depletion & Self-produced signaling molecules & Cells release and detect signals, coordinating to form multicellular structures~\cite{JOHNSON19924600}. \\
\hline
Cancer cells & Tumor microenvironment & Growth and survival signals & Directed invasion of tissues and movement toward blood vessels during metastasis~\cite{Teicher_2010}. \\
\hline
Mammalian sperm cells & Reproductive tract & Egg-derived chemical cues & Navigation toward the egg via modulation of flagellar movement~\cite{Publicover2007}. \\
\hline
Fibroblasts & Wound healing sites & Growth and repair signals & Migration into wounds to repair tissue and remodel the extracellular matrix~\cite{Barrientos_2008}. \\
\hline
Endothelial cells & Developing tissues or tumor environment & Blood vessel growth signals & Migration toward growth cues to form new blood vessels~\cite{Ferrara2005}. \\
\hline
Neuronal growth cones\footnotemark & Developing neural tissues & Guidance cues (attractants or repellents) & Extension of axons guided by attractive or repulsive signals to form neural circuits~\cite{Donnell_2009}. \\
\hline
Macrophages & Inflamed or damaged tissues & Immune signals from damaged cells & Migration to injury sites to clear debris and pathogens~\cite{Wynn2013}. \\
\hline
Sea urchin sperm & Seawater & Egg-derived attractant molecules & Swimming directed by chemical cues released from the egg~\cite{Kaupp_2008}. \\
\hline
\end{tabular}
\label{tab:chemotaxis_examples}
\end{table*}

\subsection{Molecular and Cellular Mechanisms of Chemotaxis}

At the cellular level, chemotaxis can be understood as a three-stage process of \textbf{sensing}, \textbf{signal transduction}, and \textbf{actuation}, which allow cells to transform chemical gradients into directed motion. This modular view can be directly mapped onto engineered systems, with sensors acquiring inputs, processors interpreting signals, and actuators producing outputs. An overview of the individual stages of chemotaxis is provided in Fig.~\ref{fig:schematic}. 

\begin{itemize}
    \item \textbf{Sensing}: Cells detect external chemical signals through receptors on their surface. These receptors are highly sensitive and can respond even to small concentration differences across the cell. For example, bacteria cluster receptors to amplify weak signals, while larger eukaryotic cells use families of receptors to detect chemical cue such as growth factors or chemokines. In both cases, the outcome is the ability to establish a sense of direction in a noisy environment~\cite{Kim2002}.
    
    \item \textbf{Signal transduction}: Once a signal is detected, intracellular networks process the information and convert it into internal polarity. In bacteria, this involves simple on–off molecular switches that regulate the rotation of flagella. In eukaryotic cells, signaling cascades amplify receptor inputs and establish front–back asymmetries by locally activating molecules that promote cytoskeletal growth~\cite{Katanaev2001}.
    
    \item \textbf{Actuation}: Finally, processed signals are converted into motion. Bacteria achieve this through alternating straight runs and random reorientations (tumbles), allowing them to bias movement toward favorable conditions. Eukaryotic cells instead remodel their internal skeleton: protruding their front, contracting their rear, and effectively crawling toward the source. In both systems, chemical information is ultimately translated into mechanical force and directed migration~\cite{Sourjik2012}.
\end{itemize}

\subsection{Structural and Biophysical Approaches to Chemotaxis}

Understanding chemotaxis requires insight into how cells are physically organized and how their molecular machinery behaves dynamically. Structural and biophysical approaches reveal how spatial arrangement, conformational changes, and mechanical forces shape the sensitivity and precision of chemotactic signaling. These methods help to quantify how cells transform chemical inputs into mechanical outputs within the limits imposed by their structure and biophysical environment.

\begin{itemize}    
    \item \textbf{Biophysical visualization} methods, such as advanced fluorescence imaging, track the spatiotemporal dynamics of signaling and cytoskeletal activity inside cells. They provide quantitative maps of processes like actin remodeling, calcium signaling, and membrane reorganization, offering direct observations of how information propagates through the cell~\cite{Stefan2005}.
    
    \item \textbf{Microfluidic platforms} generate controlled chemical gradients and physical environments, making it possible to measure chemotactic behavior under reproducible conditions. These devices reveal how cells adapt to changing gradients, maintain directional persistence, or switch strategies in complex environments~\cite{Irimia2010}. Such systems bridge biology and engineering by enabling precise, quantitative tests of theoretical models.
\end{itemize}

\subsection{Modeling Approaches of Chemotaxis}
Mathematical models are essential for understanding chemotaxis, capturing how cells sense, process, and respond to chemical gradients across multiple scales from single cell behavior to population dynamics.

\begin{itemize}
     \item \textbf{Biased random walk models} describe chemotactic navigation as a stochastic movement strategy in which random motion is modulated by environmental cues to generate directed migration. Unlike purely diffusive movement, in a biased random walk the probability of motion in certain directions depends on the local or temporal concentration of chemoattractants. Biased random walk models capture chemotaxis mechanism using stochastic differential equations or probabilistic rules. They scale up well when aggregated over many individuals and lead to macroscopic drift–diffusion equations such as the Patlak–Keller–Segel model~\cite{KELLER1971225}. Extensions of biased random walk model include correlated random walks with directional memory, and internal state models that incorporate intracellular signaling dynamics. These improvements allow for the simulation of more complex phenomena such as noise filtering, collective chemotaxis, and dynamic adaptation in changing environments~\cite{Freingruber2024}.

    \item \textbf{Cell density models} formulated as \acp{PDE}, describe how the density of cell populations evolves in response to chemical cues. The classical Patlak–Keller–Segel model represents the cell flux as the sum of a diffusive component and a chemotactic drift component proportional to the chemoattractant gradient~\cite{KELLER1971225,Arumugam2020}. These models can reproduce a wide range of emergent phenomena such as aggregation, pattern formation, and wave propagation across biological systems including bacterial colonies, slime molds, immune cell recruitment, and tumor invasion~\cite{PAINTER2019162}. More recent work further generalizes the model to include multiple chemical species and complex nonlinear couplings, enabling quantitative matching of experimentally observed spatial distributions and pattern dynamics~\cite{Tu2024}. These \ac{PDE} based density models are especially suited to large scale systems.
\footnotetext{Neuronal growth cones are motile extensions of neurons, not independent cells, but exhibit chemotaxis to guide axon growth.}
    \item \textbf{Mechanochemical models} couple intracellular signaling with physical processes such as shape changes, force generation, and interactions with the surrounding environment to explain how cells convert chemical gradients into directed motion~\cite{Tao2020,Feng2018,Zhou2021}. These models often use systems of \acp{ODE} or \acp{PDE} to describe the spatiotemporal evolution of a few key variables that represent internal polarity, force production, and adhesion. External signals bias these variables so that the cell develops a front and rear, enabling persistent movement~\cite{Salvadori2024}. By explicitly linking signaling and mechanics, mechanochemical models reproduce emergent behaviors such as spontaneous symmetry breaking, oscillatory movement, and long-term directional migration~\cite{Sun2017}. Such frameworks are useful not only for understanding natural processes like development or wound healing, but also for designing synthetic systems that aim to control synthetic cell-like motility~\cite{Biswas2021}.

    \item \textbf{Agent-based models} represent individual cells as autonomous agents governed by mechanistic or rule based behaviors, enabling simulation of chemotactic navigation through local sensing, signal processing, and motility responses~\cite{Emily2019, Nagarajan2022}. Unlike continuum models, which treat cells as a continuous medium and average out individual variations, agent-based models simulate individual cells to capture spatial and temporal heterogeneity, variability, stochasticity, and local interactions. This approach enables the emergence of collective patterns, such as swarming, band formation, and clustering, as described in~\cite{Witkowski2016}. This bottom-up approach also incorporates internal states (e.g., adaptation or gene regulation), enabling representation of feedback, memory, and plasticity in chemotactic behavior~\cite{Pinheiro2016}. Recent applications extend to synthetic biology, where agent-based models guide the design of engineered microbial consortia and multicellular systems, and to robotics, where they inspire communication and coordination strategies in micro-robotic swarms~\cite{Miller2010,Elgeti_2015}.
\end{itemize}

\subsection{Information-Theoretic Studies of Chemotaxis}
Information theory provides a powerful framework for studying chemotaxis as an information processing system, where cells are modeled as receivers decoding noisy chemical signals to guide motility decisions. Information-theoretic studies quantify how cells extract information from fluctuating environments, revealing the physical and biological limits of gradient sensing and informing applications in synthetic biology, \ac{MC}, and ecological modeling.

Research in this area applies concepts like mutual information, rate distortion, entropy, and physical sensing limits to understand how cells process chemical signals. For instance, authors in~\cite{Bialek2008} analyzed the information capacity of biochemical signaling cascades and showed how mutual information can be used to evaluate the fidelity of chemotactic signal transmission. In~\cite{Foster2021}, authors investigated the effect of cell to cell communication on precision of gradient sensing by developing a model that includes nearest-neighbor communication mechanisms between cells. In~\cite{Julian2024}, authors used information theory to compare the spatial and temporal strategies in chemotaxis processes for an ideal agent (cell) with unlimited information processing capabilities. The authors in~\cite{Lukas2024} explored the framework of semantic information to analyze bacterial chemotaxis. They showed how the information between the environment and the bacteria is related to the adaptation of the bacteria to the dynamic environment in order to ensure survival. In~\cite{Andrews2007,Andrews2007_Conf}, authors applied \ac{RDT} to analyze performance-cost tradeoffs in cellular decision making. The rate distortion function establishes a lower bound on the information transmission rate required through the cellular signaling network to meet a specified performance level.
Gradient sensing is fundamentally constrained by physical limits, including the number of receptors, diffusion dynamics, and the size of the cell~\cite{Berg1977,HuBo2010}. These principles guide the design of bio-inspired systems, such as synthetic molecular sensors and nanorobots for targeted drug delivery or environmental sensing, by mimicking cellular strategies for efficient signal processing~\cite{Endres2008, Kong2023}.


\begin{figure}[t]
    \centering
    \includegraphics[width=0.9\linewidth]{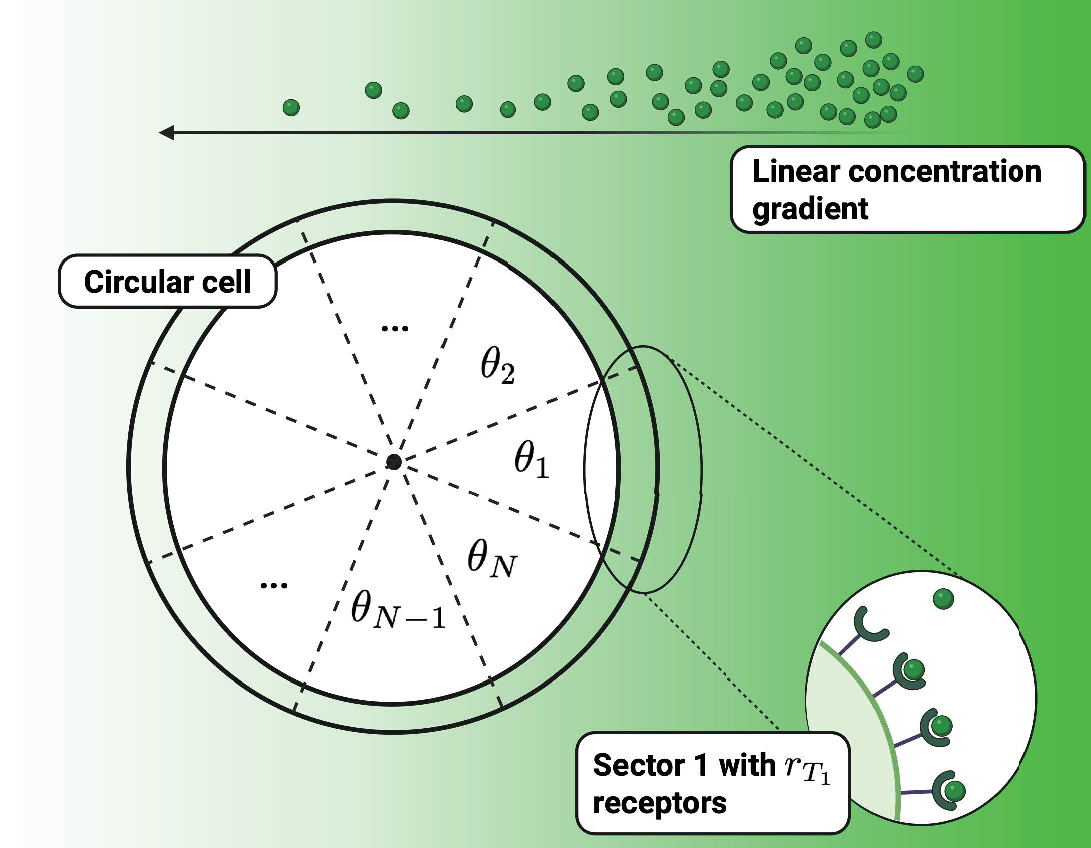}
    \caption{\small Schematic of a circular cell divided into $N$ sectors, with receptors uniformly distributed along the membrane. The chemoattractant (ligand) source generates a linear ligand gradient, visualized by a green background gradient and matching outer arc colors (darker shades indicate higher ligand concentration) (created with biorender.com).}
    \label{fig:cell_model}
\end{figure}


\section{System Model}\label{sec:system_model}

Understanding how cells sense and interpret external chemoattractant gradients to guide migration is a central question in cell biology. Extensive research has shown that cells, such as Dictyostelium discoideum and mammalian neutrophils, detect these gradients via the \ac{LEGI} mechanism~\cite{Ma2004,Azadeh2006}. In this section, we present a mathematical model of chemotaxis in Dictyostelium discoideum, based on the \ac{LEGI} mechanism, focusing on the stochastic binding of ligand-receptor complexes based on the models in~\cite{Levchenko2002,Ma2004,Andrews2007}. A schematic of the considered system model is shown in Fig.~\ref{fig:cell_model}. Assuming a circular cell geometry in 2D space and a linear ligand gradient, we first model the distribution of bound ligand-receptor complexes on the cell membrane. Then, we characterize the conditional probability distribution of the cell's movement direction given the ligand gradient direction, as derived from the \ac{LEGI} model.

We model a Dictyostelium discoideum cell as a circular object with receptors uniformly distributed over its membrane. The membrane is discretized into $N$ equal sectors, each defined by an angle $\theta_i$, where $i \in \{1, \dots, N\}$ (see Fig.~\ref{fig:cell_model}). There are $r_{\mathrm{T}_i}$ receptors in sector $i$, which independently bind ligand molecules in their local environment. A chemoattractant source located at angle $\theta_s \in \{\theta_1, \dots, \theta_N\}$ relative to the cell generates a linear ligand gradient across the space. The ligand concentration at sector $\theta_i$ is modeled as follows
\begin{align} \label{eq:lig_conc}
    l(\theta_i,\theta_s) = a - b \left(1 - \cos(\theta_i - \theta_s)\right),
\end{align}
where $a$ is the maximum ligand concentration experienced by the cell, and $b$ quantifies the gradient strength. The cosine term in~\eqref{eq:lig_conc} captures the angular dependence of the gradient, reflecting spatial variation in ligand concentration relative to the source located at $\theta_s$. 

The number of bound ligand-receptor complexes in sector $i$, denoted by $C_i$, is random due to the stochastic nature of binding. Since each receptor binds independently, the binding process can be modeled as a sequence of Bernoulli trials. With $r_{\mathrm{T}_i}$ receptors in sector $i$, the number of bound complexes follows a Binomial distribution.

To characterize the occupancy probability $f_i(\theta_s)$ of an individual receptor, we model ligand-receptor binding at steady state. The binding reaction between free receptors and ligands in sector $i$ follows as
\begin{align}
    r_i + l_i \xrightleftharpoons[k_{\text{off}}]{k_{\text{on}}} c_i,
\end{align}
where $r_i$ is a free receptor, $l_i$ a ligand, $c_i$ a bound complex, and $k_{\text{on}}$, $k_{\text{off}}$ are the association and dissociation rate constants of the receptor, respectively. At steady state, the rates balance as
\begin{align}\label{eq:steady_state}
    k_{\text{on}} \,\rho_r \,[l_i] = k_{\text{off}} \,\rho_c,
\end{align}
where $\rho_r$ and $\rho_c$ are the surface densities of free and bound receptors, and $[l_i] = l(\theta_i,\theta_s)$ is the ligand concentration in sector $i$. Solving~\eqref{eq:steady_state} yields
\begin{align}\label{eq:eq_exp}
    \rho_c = \frac{\rho_r\, l(\theta_i,\theta_s)}{K_\mathrm{d}},
\end{align}
where $K_\mathrm{d} = \tfrac{k_{\text{off}}}{k_{\text{on}}}$ is the dissociation constant, assumed uniform across all sectors~\cite{Lauffenburger1993}. The total receptor density per sector, $\rho_{\mathrm{T}_i}$, is conserved as
\begin{align}
    \rho_{\mathrm{T}_i} = \rho_r + \rho_c,
    \label{eq:recepDens}
\end{align}
with $r_{\mathrm{T}_i} = \rho_{\mathrm{T}_i} A_i$ receptors per sector~\cite{JOHNSON19924600,Masahiro2001}, where $A_i$ is the membrane area of sector $i$. 
Inserting \eqref{eq:recepDens} into~\eqref{eq:eq_exp} and solving for $\rho_c$, the density of bound receptors can be obtained as follows
\begin{align}
    \rho_c = \frac{\rho_{\mathrm{T}_i}\, l(\theta_i,\theta_s)}{K_\mathrm{d} + l(\theta_i,\theta_s)}.
\end{align}
The number of bound complexes in sector $i$ is $c_i = \rho_c A_i$, which simplifies to
\begin{align}
    c_i = \frac{r_{\mathrm{T}_i}\, l(\theta_i,\theta_s)}{K_\mathrm{d} + l(\theta_i,\theta_s)}.
\end{align}
Hence, the single receptor occupancy probability, i.e., the fraction of bound receptors in sector $i$ follows as
\begin{align}\label{eq:fi_def}
    f_i(\theta_s) = \frac{c_i}{r_{\mathrm{T}_i}}
    = \frac{l(\theta_i,\theta_s)}{K_\mathrm{d} + l(\theta_i,\theta_s)}\,.
\end{align}
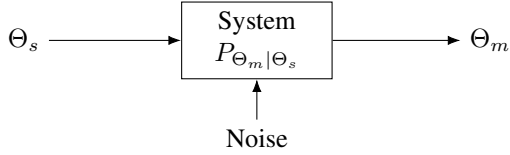
\begin{figure}[t]
    \centering
    \begin{tikzpicture}[node distance=2cm and 2cm, 
        >=Latex, 
        block/.style={draw, minimum width=2cm, minimum height=1cm, align=center},
        sum/.style={draw, circle, inner sep=1pt, minimum size=0.5cm, node contents={$+$}}]
        
        \node (stimulus) at (0,0) {$\Theta_s$};
        \node (noise) at (3.1, -1.3) {Noise};
        
        \node[block] (sensing) at (3.1,0) {System \\ $P_{\Theta_m\mid \Theta_s}$};
        
        \node (output) [right=1.7cm of sensing] {$\Theta_m$};
        
        \draw[->] (stimulus.east) -- (sensing.west);
        \draw[->] (noise) -- (sensing.south);
        \draw[->] (sensing.east) -- (output);
        
    \end{tikzpicture}
    \caption{\small Schematic of chemotaxis as an information processing system. The true source direction $\theta_s$ passes through the noisy system to yield receptor occupancy, which the strategy stage maps to a movement direction $\theta_m$.}
    \label{fig:system_flow}
\end{figure}
As receptors bind ligands independently, the number of ligand--receptor complexes in each sector follows a Binomial distribution. Specifically, the number of trials corresponds to the total number of receptors in sector $i$, $r_{\mathrm{T}_i}$, and the success probability is given by the binding probability $f_i(\theta_s)$. Hence, the number of bound receptors in sector $i$ is
\begin{align}
    c_i \sim \mathrm{Binomial}\!\left(r_{\mathrm{T}_i}, f_i(\theta_s)\right)\,.
\end{align}
To link receptor occupancy with cell movement, we employ the \ac{LEGI} mechanism~\cite{Levchenko2002,Kutscher2004}. In this framework, the cell response to ligand-receptor complex in sector $\theta_i$ is determined by the ratio of a local signal to a global reference
\begin{align}
    u(\theta_i) = \frac{(c_i - \min_k c_k)^h}{\tfrac{1}{N}\sum_{j=1}^{N} (c_j - \min_k c_k)^h}\,,
    \label{eq:legi_final}
\end{align}
where $h$ is a Hill coefficient that phenomenologically accounts for amplification. The subtraction of the minimum ligand-receptor complexes among the sectors mimics experimentally observed cellular responses, which suggest negligible or no response at the rear of the cell relative to the ligand source direction~\cite{Janetopoulos2004}. Normalizing the cell response~\eqref{eq:legi_final} then yields the conditional probability of movement direction given the source direction
\begin{align}
    P_{\Theta_m|\Theta_s}(\theta_i \mid \theta_s) = \frac{u(\theta_i)}{\sum_{j=1}^{N} u(\theta_j)}\,.
\end{align}

Figure~\ref{fig:system_flow} shows the the chemotaxis process as an information processing system, where $\Theta_s$ is the input (true source direction) that can be from any direction around the cell, while $\Theta_m$ is the output (movement direction of the cell) affected by the noise. This information processing view of the chemotaxis allows us to employ techniques and tools from information theory to analyze and understand the process.

\section{Rate Distortion Theory}\label{sec:RD_Theory}

\begin{figure}[t]
    \centering
    \begin{tikzpicture}[node distance=2cm and 2cm, 
        >=Latex, 
        block/.style={draw, minimum width=2cm, minimum height=1cm, align=center},
        sum/.style={draw, circle, inner sep=1pt, minimum size=0.5cm, node contents={$+$}}]
        
        \node (stimulus) at (0,0) {$X$};
        \node (noise) at (3.1, -1.3) {Noise};
        
        \node[block] (sensing) at (3.1,0) {System \\ $P_{Y\mid X}$};
        
        \node (output) [right=1.7cm of sensing] {$ Y $};
        
        \draw[->] (stimulus.east) -- (sensing.west);
        \draw[->] (noise) -- (sensing.south);
        \draw[->] (sensing.east) -- (output);
        
    \end{tikzpicture}
    \caption{\small System schematic of a noisy channel with input $X$ and output $Y$.}
    \label{fig:signalflow_information}
\end{figure}
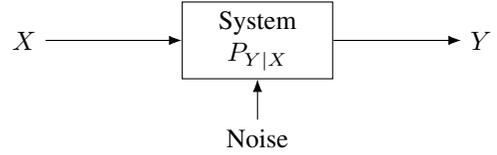

\ac{RDT} extends Shannon's information theory to lossy compression, characterizing the minimum rate of information required to represent a source while maintaining average distortion below a specified threshold~\cite{Davisson1972}. Unlike lossless compression, which ensures perfect reconstruction, \ac{RDT} permits controlled errors to reduce the information rate, mirroring biological decision-making processes like chemotaxis, where organisms balance signal fidelity with resource constraints. In the previous section, we described chemotaxis as an information-processing system in which ligand binding to cell-surface receptors initiates signaling. Within this framework, the \ac{LEGI} mechanism compares local receptor occupancy to a global reference and amplifies the response to guide the cell up the gradient. This section introduces the core concepts of \ac{RDT} for a noisy information processing system (see Fig.~\ref{fig:signalflow_information}), resembling the chemotaxis process modeled in Section~\ref{sec:system_model}. In \ac{RDT}, mutual information ($I(X;Y)$) acts as a resource, where a higher rate enables lower distortion in signal representation. In biological signaling, such as chemotaxis, higher mutual information incurs energetic costs, reflecting the trade-off between signal accuracy and resource expenditure. \ac{RDT} thus provides a framework to analyze how living systems, like Dictyostelium discoideum navigating toward the gradient source, optimize information processing under resource constraints, offering insights into efficient molecular and biological communication.

For a discrete random variable $X$ with \ac{PMF} $P_X(x)$, the entropy is defined as follows
\begin{align}
    H(X) = -\sum_{x} P_X(x) \log P_X(x),
\end{align}
quantifying the average uncertainty associated with $X$. For two random variables $X$ and $Y$, the mutual information is defined as follows
\begin{align}
\begin{split}
    I(X;Y) 
    &= H(Y) - H(Y|X) \\
    &= \sum_{x,y} P_{X,Y}(x,y) \log \frac{P_{X,Y}(x,y)}{P_X(x) P_Y(y)}\,,
\end{split}
\end{align}
measuring the reduction in uncertainty of $Y$ given $X$. In Shannon's channel coding theorem for lossless communication, the channel capacity
\begin{align}
    C = \sup_{P_X(x)} I(X;Y)\,,
\end{align}
represents the maximum rate for reliable transmission with negligible error.

In \ac{RDT}, the perfect reconstruction of an input signal is often unnecessary or infeasible. A distortion function $0 \leq d(x,y) < \infty$ quantifies the cost of representing input $x$ by an output $y$, with the choice of $d(\cdot\,,\cdot)$ depending on the application (see Table~\ref{tab:distortion_functions}).

\begin{table*}[!ht]
\scriptsize
\centering
\caption{\small Common distortion functions in \ac{RDT}.}
\label{tab:distortion_functions}
\begin{tabular}{|>{\centering\arraybackslash}m{2.9cm}|>{\centering\arraybackslash}m{3.4cm}|>{\centering\arraybackslash}m{4.4cm}|>{\centering\arraybackslash}m{5cm}|}
\hline
\textbf{Distortion Function} & \textbf{Mathematical Expression} & \textbf{Alphabet} & \textbf{Properties} \\
\hline
Hamming Distortion & 
$d(x, y) = \begin{cases} 0, & x = y, \\ 1, & x \neq y \end{cases}$ & 
Discrete source and reproduction & 
Symmetric, equal penalty for all errors, leads to closed-form solutions for binary sources. \\
\hline
Squared Distortion & 
$d(x, y) = (x - y)^2$ & 
Continuous source and reproduction in \( \mathbb{R} \) & 
Penalizes larger errors more, analytically tractable for Gaussian sources. \\
\hline
Absolute Distortion & 
$d(x, y) = |x - y|$ & 
Continuous source and reproduction in \( \mathbb{R} \) & 
Less sensitive to large errors than squared errors, linear penalty. \\
\hline
General Quadratic Distortion & 
$d(\mathbf{x}, \mathbf{y}) = (\mathbf{x} - \mathbf{y})^T \mathbf{Q} (\mathbf{x} - \mathbf{y})$ & 
Continuous source and reproduction in \( \mathbb{R}^n \) & 
Generalizes squared-error to vectors, flexible with matrix \( \mathbf{Q} \). \\
\hline
\end{tabular}
\end{table*}

For a given distortion function $d(x,y)$ and a joint probability distribution $P_{X,Y}(x,y)$, the expected distortion can be expressed as follows
\begin{align}\label{eq:ExpDist}
    \mathbb{E}[d(X,Y)] = \sum_{x,y} P_{X,Y}(x,y) \, d(x,y)\,.
\end{align}
In \ac{RDT}, the goal is to find a conditional probability distribution $P_{Y|X}(y|x)$ (decision strategy of the system) that satisfies
\begin{align}
    \mathbb{E}[d(X,Y)] \leq D\,,
\end{align}
while minimizing $I(X;Y)$ and $D$ is the distortion constraint. Considering the goal of the \ac{RDT}, the optimal decision strategy can be formulated as
\begin{align}
    P_{Y|X}^{\star}(y|x) = \operatorname*{arg\,min}_{P_{Y|X}(y|x):\,\mathbb{E}[d(X,Y)] \leq D} I(X;Y)\,.
\end{align}
The rate-distortion function is
\begin{align}
    R(D) = \min_{P_{Y|X}(y|x):\,\mathbb{E}[d(X,Y)] \leq D} I(X;Y)\,.
    \label{eq:ratedistortioncurve}
\end{align}
Rate distortion function $R(D)$ is the minimum amount of information (number of bits per symbol) that must be processed to ensure the input representation by the output is ``good enough'', where ``good enough'' is defined by the distortion constraint $D$.
Closed-form solutions for $R(D)$ exist for specific cases, such as Gaussian sources with squared-error distortion or binary sources with Hamming distortion~\cite{elementstheory1999}. In general, the rate distortion function $R(D)$ is found by solving the following optimization problem
\begin{align}
    \min_{P_{Y|X}} I(X;Y) \quad \text{s.t.} \quad \mathbb{E}[d(X,Y)] \leq D\,.
\end{align}
Using Lagrange multipliers, this is equivalent to minimizing
\begin{align}
    \mathcal{L}[P_{Y|X}] = I(X;Y) + \lambda \big( \mathbb{E}[d(X,Y)] - D \big)\,,
\end{align}
where $\lambda > 0$ is the Lagrange multiplier. The \ac{BAA}~\cite{blahut} computes $R(D)$ and $P_{Y|X}^\star$ iteratively, with inputs $P_X(x)$ and $d(x,y)$. The individual steps of \ac{BAA} are as follows:

\begin{enumerate}
    \item \textbf{Initialization:} Initialize the marginal distribution $P_Y(y)$ to a uniform distribution. This provides a starting point for the algorithm, ensuring all possible output symbols $y$ have equal probability initially.
    \item \textbf{Update $P_{Y|X}$:} For the current marginal distribution $P_Y(y)$, compute the conditional probability distribution $P_{Y|X}(y|x)$ using
    \begin{align}
        P_{Y|X}(y|x) = \frac{P_Y(y) \, \mathrm{e}^{-\lambda d(x,y)}}{\sum_{y'} P_Y(y') \, \mathrm{e}^{-\lambda d(x,y')}}\,,
        \label{eq:blahut_update}
    \end{align}
    where $\lambda > 0$ is the Lagrange multiplier and $d(x,y)$ is the distortion function. This step derives the conditional distribution that minimizes the Lagrangian $\mathcal{L}[P_{Y|X}] = I(X;Y) + \lambda \big( \mathbb{E}[d(X,Y)] - D \big)$ for the fixed $P_Y(y)$, as it satisfies the optimality condition for the rate distortion function.
    \item \textbf{Update $P_Y$:} Recompute the marginal distribution $P_Y(y)$ using the updated conditional distribution:
    \begin{align}
        P_Y(y) = \sum_x P_X(x) \, P_{Y|X}(y|x)\,.
    \end{align}
    This step ensures that $P_Y(y)$ is consistent with the joint distribution $P_{X,Y}(x,y) = P_X(x) P_{Y|X}(y|x)$, where $P_X(x)$ is the source distribution.
    \item \textbf{Iteration:} Repeat steps 2 and 3 until the distributions $P_{Y|X}(y|x)$ and $P_Y(y)$ converge.
\end{enumerate}
At convergence, the \ac{BAA} yields the optimal conditional distribution $P_{Y|X}^\star(y|x)$ for a given $\lambda$, with the mutual information $I(X;Y)$ yielding the rate-distortion function $R(D)$ and the expected distortion $\mathbb{E}[d(X,Y)] = \sum_{x,y} P_X(x) P_{Y|X}^\star(y|x) d(x,y)$. The parameter $\lambda > 0$ governs the trade-off between rate and expected distortion: larger $\lambda$ values prioritize lower distortion at the cost of higher rate, while smaller $\lambda$ values allow higher distortion for lower rate. By varying $\lambda$, the algorithm traces the rate distortion curve $R(D)$, where $\lambda$ corresponds to the negative slope $-\partial R(D)/\partial D$ at each point, producing a family of optimal strategies $P_{Y|X}^\star(y|x)$ for different distortion constraints $D$.

\section{Inverse Blahut-Arimoto Algorithm}\label{sec:IBAA}

In this section, we reformulate the problem so that, instead of using the \ac{BAA} to find the optimal decision strategy for a known distortion function, we identify the distortion function corresponding to a given decision strategy. To this end, we propose the \ac{IBAA}, which computes the distortion function $d(x,y)$ implicitly considered by the system as the criterion for developing its decision strategy. The proposed \ac{IBAA} extends the idea of distortion function computation in the binary case presented in~\cite{Porter2012}.
According to the \ac{BAA}, an optimal strategy $P^{\star}_{Y|X}(y|x)$ can be obtained via~\eqref{eq:blahut_update} for different values of the expected distortion. Given both the marginal distribution $P_Y(y)$ and the optimal strategy $P^{\star}_{Y|X}(y|x)$, it is in principle possible to compute the distortion function $d(x,y)$. However, obtaining a closed-form expression for $d(x,y)$ is generally intractable. To address this challenge, we introduce the \ac{IBAA} as a computation method for the distortion function. Since the optimal strategy satisfies~\eqref{eq:blahut_update}, we highlight key observations derived from this expression, which will serve as the foundation for the proposed \ac{IBAA}.

\begin{enumerate}
\item \label{txt:zeroArg}
If the joint probability $P_{X,Y}(x_j,y_i)$ of a particular input output pair is zero, then by Bayes' rule the corresponding conditional probability $P_{Y|X}(y_i|x_j)$ must also be zero. From~\eqref{eq:blahut_update}, setting $P_{Y|X}(y_i|x_j)=0$ requires the numerator to be zero for a fixed $\lambda$. This occurs when $P_Y(y_i)=0$, since it directly multiplies the exponential term in the numerator. In this case, the distortion value $d(x_j,y_i)\in\mathbb{R}^+$ can be assigned arbitrarily for that realization. 

Conversely, if $P_Y(y_i)\neq 0$ but $P_{Y|X}(y|x)=0$, then the exponential factor must be zero, i.e., $\mathrm{e}^{-\lambda d(x_j,y_i)}=0$, which implies $d(x_j,y_i)\to\infty$. Such infinite distortion, however, is not admissible by definition. Therefore, to ensure that $d(x,y)$ remains finite, it must hold that whenever $P_Y(y_i)>0$, the joint probability $P_{X,Y}(x_j,y_i)$ is strictly positive for every feasible event.
\item \label{txt:lamSacling} Since $d(x,y)$ appears in~\eqref{eq:blahut_update} only through the product $\lambda\, d(x,y)$, and $\lambda$ is a Lagrange multiplier varied along the rate distortion curve, the distortion function can be rescaled without altering the optimal decision strategies. Specifically, let $d'(x,y)=a\,d(x,y)$ with $a>0$, and define a new multiplier $\lambda'=\lambda/a$. Substituting into~\eqref{eq:blahut_update} yields
\begin{align}
\begin{split}
    \tilde{P}_{Y|X}(y|x) 
    &= \frac{P_Y(y)\, e^{-\lambda' d'(x,y)}}{\sum_{y'} P_Y(y')\, e^{-\lambda' d'(x,y')}} \\
    &= \frac{P_Y(y)\, e^{-\lambda d(x,y)}}{\sum_{y'} P_Y(y')\, e^{-\lambda d(x,y')}} \\
    &= P_{Y|X}(y|x)\,.
\end{split}
\end{align}
Hence, the scaling factor $a$ cancels out, implying that the strategies are invariant under positive rescaling of the distortion function.

More generally, since $P_{Y|X}(y|x)$ depends only on the product $\lambda\, d(x,y)$ together with $P_Y(y)$, any positive scaling of $d(x,y)$ can always be compensated by an appropriate adjustment of $\lambda$. Consequently, when $\lambda$ is unknown, the decision strategy does not uniquely determine the distortion function, i.e., it can be identified only except the multiplicative constant. A similar observation was made in~\cite{Porter2012}.

\item \label{txt:offsetArg} The distortion functions $d(x,y)$ and $d(x,y)+b(x,\lambda)$ lead to the same strategy for a given $P_Y(y)$
\begin{align}
\begin{split}
    P'_{Y|X}(y|x) 
    &= \frac{P_Y(y)\, e^{-\lambda \left(d(x,y)+b(x,\lambda)\right)}}{\sum_{y'} P_Y(y')\, e^{-\lambda \left(d(x,y')+b(x,\lambda)\right)}} \\[4pt]
    &= \frac{P_Y(y)\, e^{-\lambda d(x,y)}\, e^{-\lambda b(x,\lambda)}}{\Big(\sum_{y'} P_Y(y')\, e^{-\lambda d(x,y')}\Big)\, e^{-\lambda b(x,\lambda)}} \\[4pt]
    &= \frac{P_Y(y)\, e^{-\lambda d(x,y)}}{\sum_{y'} P_Y(y')\, e^{-\lambda d(x,y')}} \\[4pt]
    &= P_{Y|X}(y|x)\,.
\end{split}
\end{align}
Thus, adding an offset term $b(x,\lambda)$ to the distortion function does not affect the decision strategy $P_{Y|X}(y|x)$ or the output distribution $P_Y(y)$.

\end{enumerate}
In the following, we use Observations \ref{txt:zeroArg}, \ref{txt:lamSacling}, and \ref{txt:offsetArg} to construct the \ac{IBAA}, which provides a method for estimating the distortion function $d(x,y)$.

When collecting samples from experiments or simulations, it is possible that the absolute number of observed events for a particular input–output pair $(x_j, y_i)$ is zero, i.e., $L(x_j, y_i) = 0$. In this case, the corresponding probabilities satisfy $P_{X,Y}(x_j, y_i) = 0$ and $P_{Y|X}(y_i|x_j) = 0$, as noted in Observation~\ref{txt:zeroArg}. However, from Observation~\ref{txt:zeroArg} we also know that the situation $P_{Y|X}(y_i|x_j) = 0$ with $P_Y(y_i) \neq 0$ leads to an ill-defined distortion function $d(x_j, y_i)$, since it would require $d(x_j, y_i) \to \infty$, which is not admissible.  

To address this issue and prevent zero probabilities for feasible but rarely observed events, we apply Laplace smoothing~\cite{laplaceSmoothing}. This technique modifies the empirical distribution so that every feasible $(x, y)$ pair is assigned a non-zero probability. The smoothed joint probability is given by  
\begin{align}
P_{X,Y}(x, y) = \frac{L(x, y) + 1}{\sum_{x'} \sum_{y'} \big(L(x', y') + 1\big)} \, .
\label{eq:laplace}
\end{align}
This adjustment guarantees that the distortion function remains finite, $d(x, y) < \infty$, for all feasible input output pairs $(x_i,y_i)$. Moreover, as the number of samples increases, the influence of the smoothing diminishes, and the smoothed distribution converges to the true empirical distribution.

In the following, we derive an expression to compute $d(x, y)$ that satisfies~\eqref{eq:blahut_update}. First, note that the denominator of~\eqref{eq:blahut_update} depends only on $x$ and $\lambda$. We can therefore rewrite~\eqref{eq:blahut_update} as
\begin{align}\label{eq:22f}
    P_{Y|X}(y|x) = \frac{P_Y(y)\, e^{-\lambda d(x,y)}}{f(x, \lambda)} \, ,
\end{align}
where
\begin{align}
    f(x, \lambda) = \sum_{y'} P_Y(y') \, e^{-\lambda d(x, y')} \, .
\end{align}
Rearranging~\eqref{eq:22f} leads to
\begin{align}
    \frac{P_{Y|X}(y|x) \, f(x, \lambda)}{P_Y(y)} = e^{-\lambda d(x,y)} \, .
\end{align}
Taking the natural logarithm on both sides yields
\begin{align}
\begin{split}
    -\frac{1}{\lambda} \ln\!\left( \frac{P_{Y|X}(y|x)}{P_Y(y)} \right) 
    &= d(x,y) + \frac{\ln f(x, \lambda)}{\lambda}
    \\&= \tilde{d}(x,y) \, .
\end{split}
\end{align}
Hence, the distortion function can be determined as follows
\begin{align} \label{eq:d_xy}
    d(x,y) = \tilde{d}(x,y) - \frac{\ln f(x, \lambda)}{\lambda} \, ,
\end{align}
except the additive term $\tfrac{\ln f(x, \lambda)}{\lambda}$ depending only on $x$ and $\lambda$. Based on Observation~\ref{txt:offsetArg}, such an additive offset does not affect the optimality of the strategy, so $\tilde{d}(x,y)$ still satisfies~\eqref{eq:blahut_update}.

Next, we must choose $\lambda$. Observation~\ref{txt:lamSacling} states that $\lambda$ only scales the distortion function and does not alter the decision strategy. For simplicity, we set $\lambda = 1$, which preserves the shape of the distortion function. Inserting $\lambda = 1$ in~\eqref{eq:d_xy} and reformulating leads to
\begin{align}
    \tilde{d}(x,y) = -\ln \frac{P_{Y|X}(y|x)}{P_Y(y)} \, ,
    \label{eq:hatdist}
\end{align}
which still satisfies~\eqref{eq:blahut_update}.  

In the end, we enforce non-negativity of the distortion function. Since any offset $b(x,\lambda=1)$ can be added without violating~\eqref{eq:blahut_update}, we subtract the minimum value of $\tilde{d}(x,y)$ over $y$ to guarantee $d(x,y) \geq 0$. A valid distortion function is therefore
\begin{align}\label{eq:iblahutCore}
\begin{split}
    d(x,y) 
    &= \tilde{d}(x,y) - \min_{y} \tilde{d}(x,y) \\
    &= -\ln \frac{P_{Y|X}(y|x)}{P_Y(y)}
       - \min_{y} \left\{ -\ln \frac{P_{Y|X}(y|x)}{P_Y(y)} \right\}.
\end{split}
\end{align}

Finally, The proposed \ac{IBAA} can be summarized as follows
\begin{enumerate}
    \label{txt:IBAASummarized}
    \item Estimate the joint distribution \(P_{X,Y}(x,y)\) from experiments or simulations using Laplace smoothing in~\eqref{eq:laplace}.
    \item Compute the marginal and conditional distributions
    \begin{align}
        P_Y(y) &= \sum_{x'} P_{X,Y}(x',y)\,, \\
        P_{Y|X}(y|x) &= \frac{P_{X,Y}(x,y)}{\sum_{y'} P_{X,Y}(x,y')}\,.
    \end{align}
    \item Set \(\lambda = 1\) and compute \(\tilde{d}(x,y)\) using~\eqref{eq:hatdist}.
    \item Compute $d(x,y)$ by~\eqref{eq:iblahutCore} 
    \begin{align}
        d(x,y) = \tilde{d}(x,y) - \min_{y} \tilde{d}(x,y)\,.
    \end{align}
\end{enumerate}
This procedure yields a feasible distortion function $d(x,y)$ that satisfies~\eqref{eq:blahut_update}. In the context of analyzing information processing systems such as chemotaxis, \ac{IBAA} allows us to find out the shape of the distortion function $d(x,y)$ that the system considered for decision making.

\section{Evaluation of the Proposed \ac{IBAA}}\label{sec:results}

In this section, we evaluate the proposed \ac{IBAA} for computing the distortion function. We first construct a binary apoptosis model based on predefined theoretical distortion functions. Then, we apply the proposed \ac{IBAA} to this model to estimate the corresponding distortion functions and compare them with the theoretical ones. Next, we apply \ac{IBAA} to the \ac{LEGI} model derived in Section~\ref{sec:system_model}, to compute the distortion function for a chemotaxis process.

\subsection{Apoptosis}\label{sec:binary_case}
We model apoptosis, or programmed cell death, as a binary decision process critical for development, immune function, and cancer prevention, where dysregulated cell death or survival can lead to pathologies~\cite{Porter2012}. In this model, the input $X$, representing the number of initiator \ac{c8} molecules, determines the binary output $Y$, where $y=1$ indicates apoptosis (cell death) and $y=0$ indicates survival. A threshold mechanism governs the decision: if the number of \ac{c8} molecules $x$ exceeds a critical threshold $x_{\text{th}}$, downstream pathways trigger apoptosis; otherwise, the cell remains viable. Due to the stochastic nature of molecular interactions and low \ac{c8} counts, a deterministic mapping from $X$ to $Y$ is infeasible. Instead, we use a probabilistic model with the conditional distribution $P_{Y|X}(y|x)$, aligned with \ac{RDT} to capture the trade-off between signal fidelity and noise in cellular decision-making. For the distribution of \ac{c8} molecules $P_X(x)$ inside the cell, we adopt an exponential distribution, following the modeling approach in~\cite{Porter2012} 
\begin{align}
	P_X(x)= \begin{cases}
		\gamma e^{-\gamma x}, & \text{if } x \geq 0\,, \\
		0, & \text{else}\,,
	\end{cases}\label{eq:exp_dist}
\end{align}
with the rate parameter $\gamma = 0.5$. This distribution reflects the higher likelihood of fewer \ac{c8} molecules and the increasing rarity of larger counts.

We analyze the impact of two distortion functions, Hamming-like~\eqref{eq:Hamming} and rectified squared-error~\eqref{eq:squared}, on the decision strategy for cellular apoptosis, with a threshold $x_\text{th}=600$. The Hamming-like distortion function is defined as
\begin{align}
d(x,y) = \begin{cases}
\begin{array}{c|c|c}
     & y=0 & y=1 \\ \hline
x < x_\text{th} & 0 & 1 \\ \hline
x \geq x_\text{th} & 1 & 0
\end{array}
\end{cases}.
\label{eq:Hamming}
\end{align}
The rectified squared distortion function, closely matched with the empirical apoptosis distortion function observed in~\cite{Porter2012}, yields $0$ for correct decisions and a squared error otherwise
\begin{align}
d(x,y) = \begin{cases}
\begin{array}{c|c|c}
     & y=0 & y=1 \\ \hline
x < x_\text{th} & 0 & \frac{(x - x_\text{th})^2}{20000} \\ \hline
x \geq x_\text{th} & \frac{(x - x_\text{th})^2}{20000} & 0
\end{array}
\end{cases}.
\label{eq:squared}
\end{align}

\begin{figure}[t]
    \centering
    \begin{minipage}{0.49\linewidth}
        \centering
            \includegraphics[width=\linewidth]{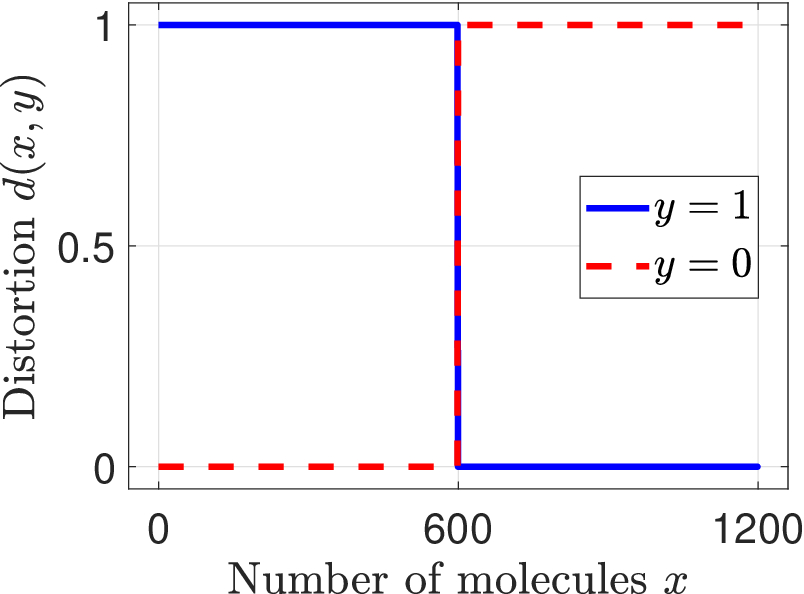}
        \\[0.3em] 
        (a)
    \end{minipage}
    \hfill
    \begin{minipage}{0.49\linewidth}
        \centering
        \includegraphics[width=\linewidth]{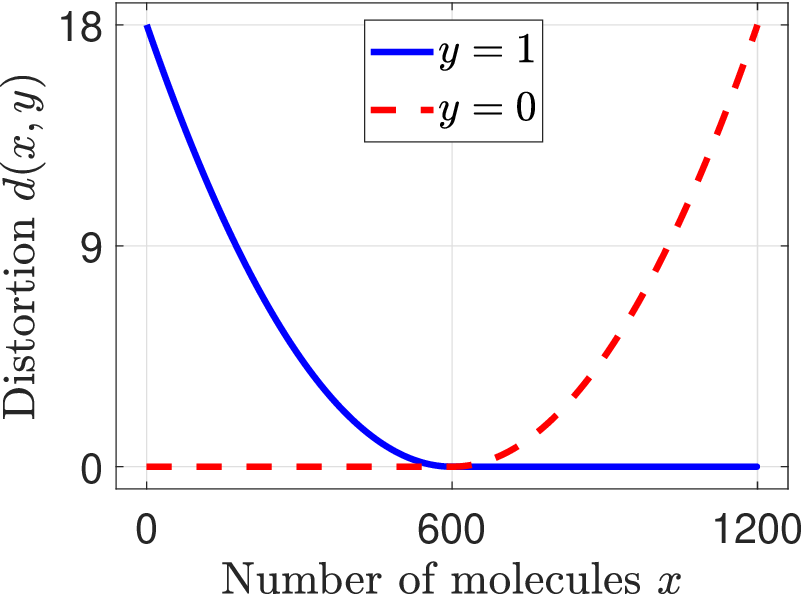}
        \\[0.3em]
        (b)
    \end{minipage}
    \caption{\small (a) Hamming-like distortion function in~\eqref{eq:Hamming}; 
    (b) Rectified squared distortion function in~\eqref{eq:squared}.}
    \label{fig:originalD_binary}
\end{figure}

To verify the proposed \ac{IBAA} in Section~\ref{sec:IBAA}, we first obtain the optimal strategy using the \ac{BAA} from Section~\ref{sec:RD_Theory}. Using the input–output data generated by this optimal strategy, we then apply \ac{IBAA} to estimate the distortion function. The estimated distortion is compared with the theoretical distortion functions defined in~\eqref{eq:Hamming} and~\eqref{eq:squared}. Figure~\ref{fig:originalD_binary} shows the theoretical distortion functions in~\eqref{eq:Hamming} and~\eqref{eq:squared}. The blue (red) curve indicates the distortion for $y=1$ ($y=0$). In Fig.~\ref{fig:originalD_binary}a, the Hamming-like distortion yields $d(x,y)=0$ when $x \geq x_{\text{th}}$ and $y=1$ or $x < x_{\text{th}}$ and $y=0$, and $d(x,y)=1$ otherwise. Figure~\ref{fig:originalD_binary}b shows the rectified squared distortion function in~\eqref{eq:squared}, were in case of mismatch between the intended input message and the output, $d(x,y) = \frac{(x - x_{\text{th}})^2}{20000}$, increasing with the deviation of $x$ from $x_{\text{th}}$.

\begin{figure}[t]
    \centering
    \includegraphics[width=\linewidth]{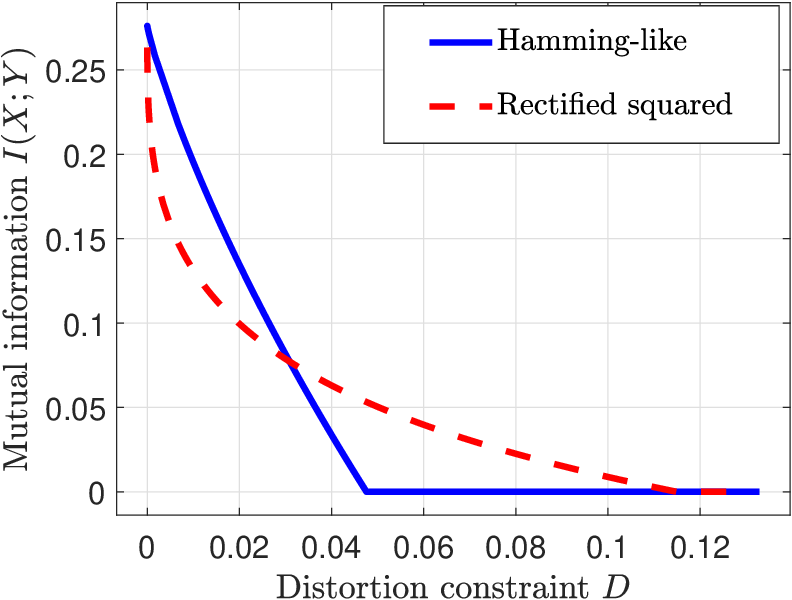}
    \caption{\small Rate distortion curves $R(D)$ for the Hamming-like distortion function~\eqref{eq:squared} (blue), and the rectified squared distortion function~\eqref{eq:Hamming} (red).}
    \label{fig:ratecurve_binary}
\end{figure}

With \ac{BAA} we can calculate optimum strategies corresponding to different levels of distortion and the rate distortion curve~\eqref{eq:ratedistortioncurve} by varying the Lagrange multiplier $\lambda$. The rate distortion curves for both distortion functions~\eqref{eq:Hamming} and~\eqref{eq:squared} are shown in Fig~\ref{fig:ratecurve_binary}. The rate distortion curve describes the minimum amount of mutual information (unit of bits) between stimulus and response required for the cell to achieve a given expected distortion ($\mathbb{E}[d(X,Y)] \leq D$) when making an apoptosis decision. From Fig.~\ref{fig:ratecurve_binary}, it can be observed that for both considered distortion functions, the rate distortion is a decreasing function of the required distortion constraint. We can observe that in case of Hamming-like and rectified squared distortion function the mutual information $I(X;Y)$ between input $X$ and output $Y$ of the system is zero at $D=0.048$ and $D=0.115$, respectively. This indicates that the output is chosen independently of the input, implying no information is transmitted from input to output. Furthermore, for a fixed distortion constraint $D$, a higher rate distortion implies that more mutual information is required to meet the specified distortion criterion.

\begin{figure}[ht]
    \centering
    \begin{minipage}{0.49\linewidth}
        \centering
        \includegraphics[width=\linewidth]{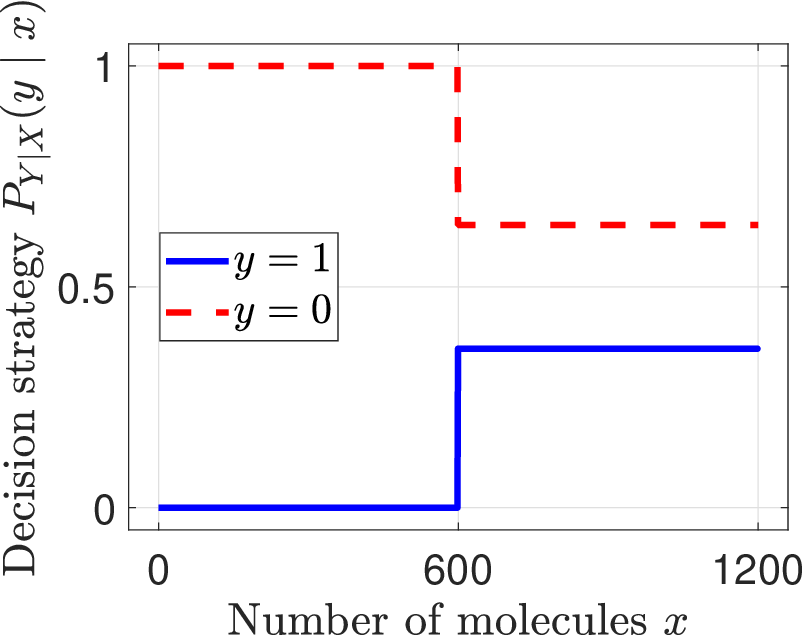}
        \\[0.3em]
        (a)
        \label{fig:binarycaseHamminglow}
    \end{minipage}
    \hfill
    \begin{minipage}{0.49\linewidth}
        \centering
        \includegraphics[width=\linewidth]{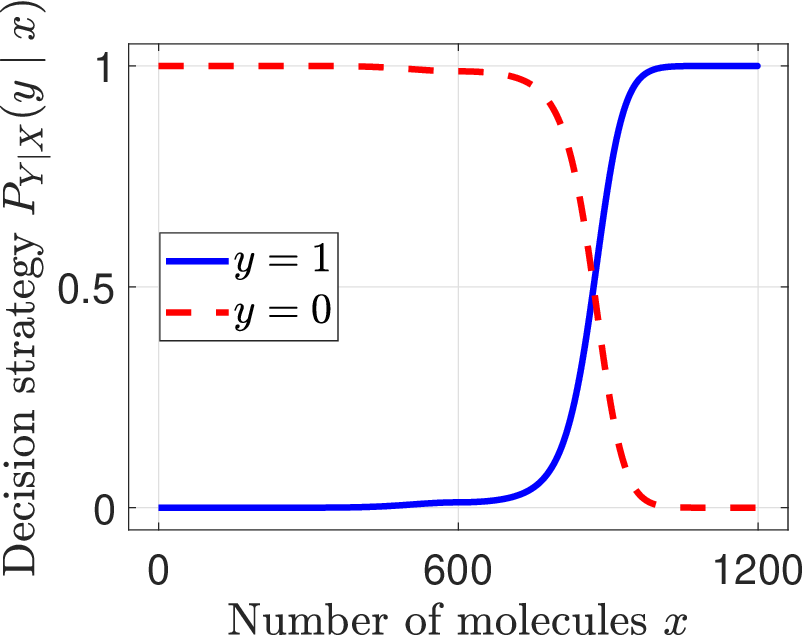}
        \\[0.3em]
        (b)
        \label{fig:binarycasesquaredlow}
    \end{minipage}

    \vspace{1em}

    \begin{minipage}{0.49\linewidth}
        \centering
        \includegraphics[width=\linewidth]{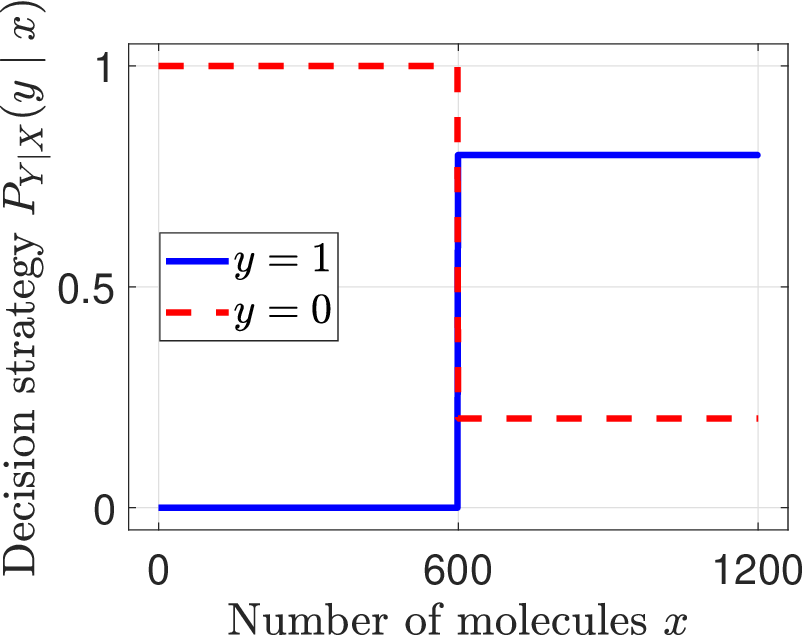}
        \\[0.3em]
        (c)
        \label{fig:binarycaseHamminghigh}
    \end{minipage}
    \hfill
    \begin{minipage}{0.49\linewidth}
        \centering
        \includegraphics[width=\linewidth]{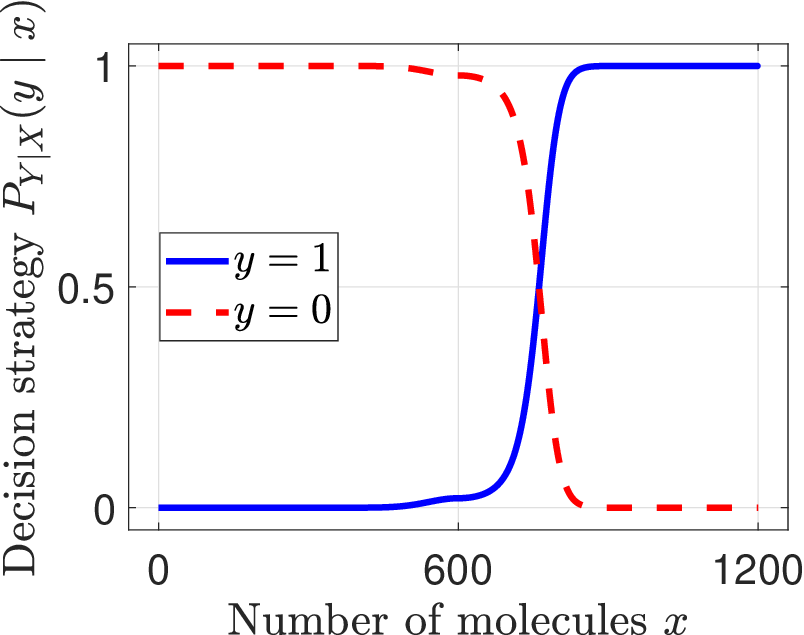}
        \\[0.3em]
        (d)
        \label{fig:binarycasesquaredhigh}
    \end{minipage}

    \caption{\small Decision strategies $P_{Y|X}(y|x)$ for different distortion functions: 
    (a) Hamming-like distortion with $D=0.031$ and $R(D)=0.077$, 
    (b) Rectified squared distortion with $D=0.031$ and $R(D)=0.077$, 
    (c) Hamming-like distortion with $D=0.01$ and $R(D)=0.2$, 
    (d) Rectified squared distortion with $D=0.01$ and $R(D)=0.13$. 
    }
    \label{fig:strategies_binary}
\end{figure}

Figure~\ref{fig:strategies_binary} illustrates the decision strategies $P_{Y\mid X}(y\mid x)$ for different distortion constraints. 
Figures~\ref{fig:strategies_binary}a and \ref{fig:strategies_binary}c show the optimal strategies $P_{Y\mid X}(y|x)$ for a Hamming-like distortion function at $D=0.031$ and $D=0.01$, respectively. 
For $D=0.031$, the decision strategy can be interpreted as follows: for input values below the threshold $x_{\text{th}}$, the output is reconstructed deterministically from the sensed input, while for inputs above $x_{\text{th}}$ the output is chosen probabilistically, with probability $0.64$ for $y=1$ and $0.36$ for $y=0$. This behavior stems from the source probability distribution~\eqref{eq:exp_dist}. Since $x$ follows an exponential distribution, smaller input values (below the threshold) are more likely, and the system effectively exploits this prior knowledge by introducing probabilistic decisions for large inputs when the distortion constraint is higher. As a result, the decision strategy becomes asymmetric: outputs for large $x$ are less certain than for small $x$. 
In contrast, at lower distortion ($D=0.01$), Figs~\ref{fig:strategies_binary}c and~\ref{fig:strategies_binary}d, when mutual information is high, the strategy becomes nearly deterministic and directly tracks the input, with little influence from the prior distribution.

Figures~\ref{fig:strategies_binary}b and \ref{fig:strategies_binary}d show the optimal decision strategies for the rectified squared distortion function for $D=0.031$, where the mutual information equals that of the Hamming-like strategy according to Fig.~\ref{fig:ratecurve_binary}, and $D=0.01$. Notably, for the same distortion constraints, the strategies under the rectified squared distortion function differ from those under the Hamming-like distortion function, emphasizing the impact of the distortion function's structure on the resulting decision strategies.
For $D=0.031$, even for inputs above the threshold $x_{\mathrm{th}}$, the probability of $y=1$ (apoptosis) remains low or near zero. This reflects the influence of the prior distribution: since large $x$ values are rare under the exponential input, the system minimizes distortion by relying more on prior knowledge than on the sensed signal. 
At lower distortion ($D=0.01$), the strategy shifts leftward, becoming more sensitive to the input and less dependent on prior knowledge, which is the input distribution.

\begin{figure}[ht]
    \centering
    \begin{minipage}{0.49\linewidth}
        \centering
        \includegraphics[width=\linewidth]{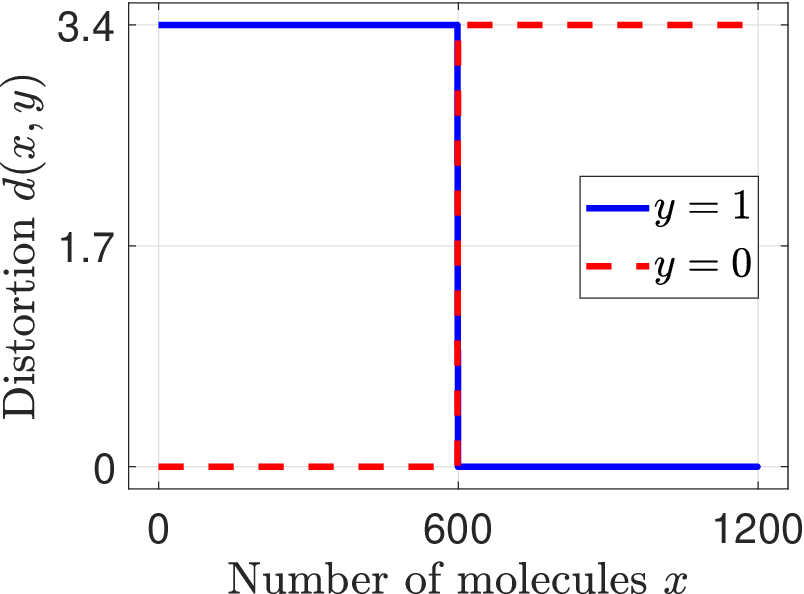}
        \\[0.3em]
        (a)
        \label{fig:binarydistortionHamminglow}
    \end{minipage}
    \hfill
    \begin{minipage}{0.49\linewidth}
        \centering
        \includegraphics[width=\linewidth]{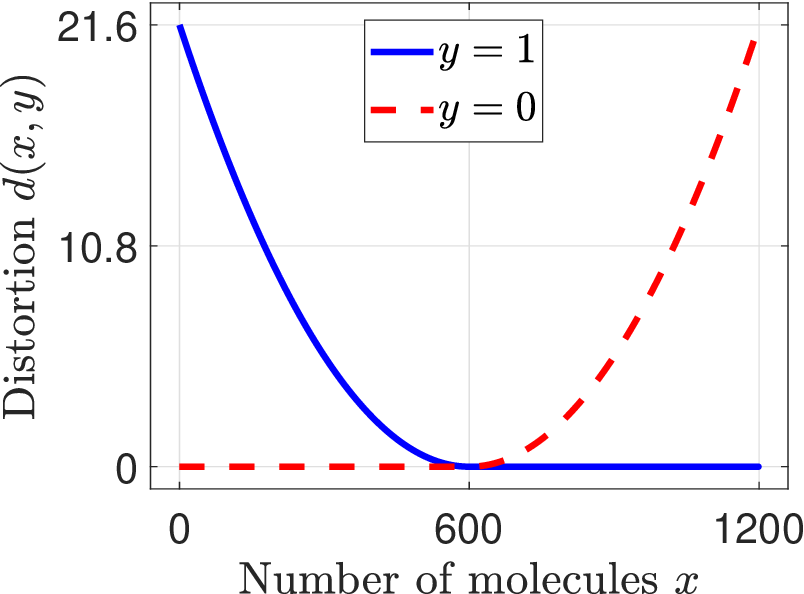}
        \\[0.3em]
        (b)
        \label{fig:binarydistortionsquaredlow}
    \end{minipage}

    \vspace{1em}

    \begin{minipage}{0.49\linewidth}
        \centering
        \includegraphics[width=\linewidth]{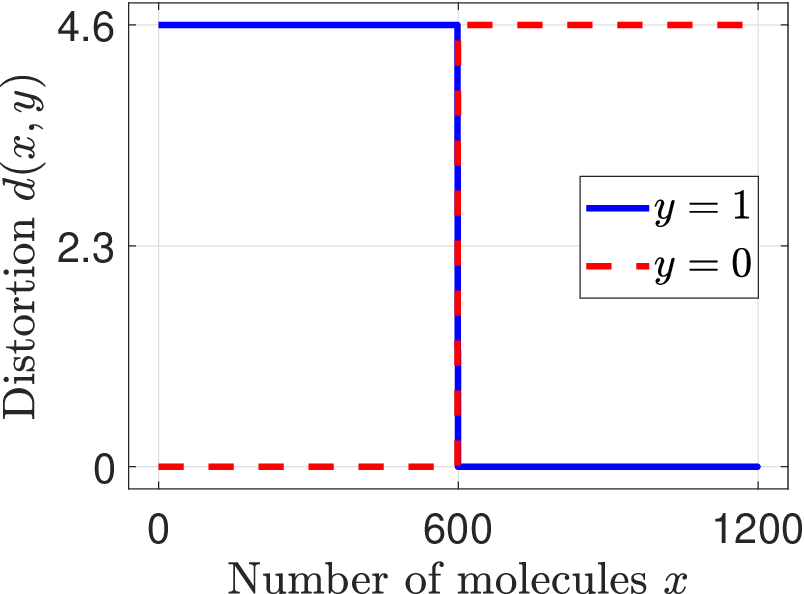}
        \\[0.3em]
        (c)
        \label{fig:binarydistortionHamminghigh}
    \end{minipage}
    \hfill
    \begin{minipage}{0.49\linewidth}
        \centering
        \includegraphics[width=\linewidth]{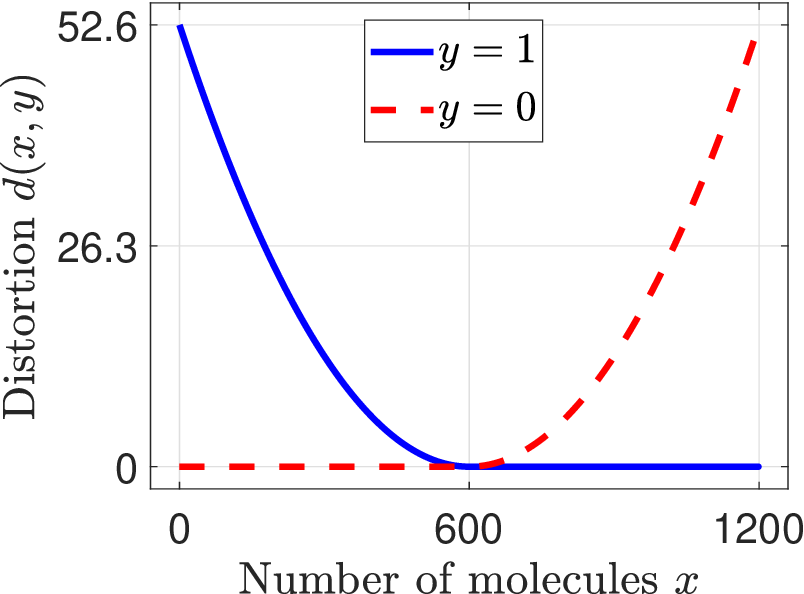}
        \\[0.3em]
        (d)
        \label{fig:binarydistortionsquaredhigh}
    \end{minipage}

    \caption{\small Estimated distortion functions obtained with \ac{IBAA}: 
    (a) Hamming-like distortion with $\lambda=3.44$, 
    (b) Rectified squared distortion with $\lambda=1.2$, 
    (c) Hamming-like distortion with $\lambda=4.6$, 
    (d) Rectified squared distortion with $\lambda=2.93$.}
    \label{fig:distortionscalc_binary}
\end{figure}

To evaluate the proposed \ac{IBAA} and compute the distortion function from the system's perspective, we use the decision strategy $P_{Y|X}(y|x)$ and the output distribution $P_Y(y)$ from the simulation data. The output distribution is obtained using the input distribution and the decision strategy based on the Bayes' rules. Figure~\ref{fig:distortionscalc_binary} shows the computed distortion functions for different values of $\lambda$. We observe that \ac{IBAA} accurately estimates the shape of the distortion functions, with $\lambda$ acting as a scaling factor. The differences between the distortion functions are expressed only as a scaling factor, as expected.

\subsection{Chemotaxis}
In this section, we apply the proposed \ac{IBAA} (introduced in Section~\ref{sec:RD_Theory}) to compute the distortion function associated with chemotaxis. Specifically, we use the \ac{LEGI} model from Section~\ref{sec:system_model} as the underlying representation of cellular gradient sensing. All physical parameters employed in the chemotaxis simulation were taken from~\cite{Andrews2007} and reported in Table~\ref{tab:param}.

\begin{figure}[t]
	\centering
\begin{tikzpicture}
    \def\thetax{30}   
    \def\thetay{120}  

    \coordinate (O) at (0,0);

    \draw (O) circle (0.3cm);

    \coordinate (Px) at ({0.3 * cos(\thetax)}, {0.3 * sin(\thetax)});
    \coordinate (Py) at ({0.3 * cos(\thetay)}, {0.3 * sin(\thetay)});

    \coordinate (Ex) at ($(Px) + ({2*cos(\thetax)}, {2*sin(\thetax)})$);
    \coordinate (Ey) at ($(Py) + ({2*cos(\thetay)}, {2*sin(\thetay)})$);

    \draw[thick, red, ->] (Px) -- (Ex);
    \draw[thick, blue, ->] (Py) -- (Ey);

    \node[red, anchor=west] at (Ex) {source direction};
    \node[blue, anchor=west] at (Ey) {\,movement direction};

    \node[below=8pt] at (O) {\textbf{cell}};

    \draw[thick] (-1,0) -- (3,0) coordinate (Xaxis);

    \pic [draw=red, ->, "$\theta_s$", angle eccentricity=1.2, angle radius=10mm]
        {angle=Xaxis--O--Px};
    \pic [draw=blue, ->, "$\theta_m$", angle eccentricity=1.2, angle radius=15mm]
        {angle=Xaxis--O--Py};
\end{tikzpicture}

\caption{\small Schematic of a cell directional information processing. The red arrow indicates the direction of the ligand source and the blue blue arrow shows the direction selected by the cell according.}
\label{fig:anglesOfCell}
\end{figure}
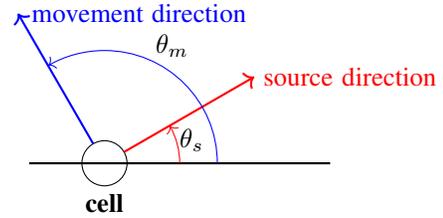
\begin{table}[!t]
\begin{center}
\caption{\small LEGI model parameters.}\label{tab:param}
\resizebox{1\columnwidth}{!}{
 \begin{tabular}{|| c | c | c ||}
 \hline
 Variable & Definition & Value \\ [0.5ex] 
 \hline\hline
  $a$ & Maximum ligand concentration experienced by the cell & $220$ \\ 
  \hline
 $b$ & Gradient strength & $20$\\ 
 \hline
 $K_\mathrm{d}$ & Dissociation constant & $200$\\
 \hline
 $r_{\mathrm{T}_i}$ & Total receptors per sector,  & $1000$ \\
 \hline
   $N$ & Number of sectors along the cell membrane & $100$ \\
 \hline
\end{tabular}}
\end{center}
\end{table}
We assume that the cellular decision strategy is optimal in nature. The ligand concentration arrives towards the cell from the direction $\theta_s$. Based on this concentration profile, the \ac{LEGI} model determines the decision for the movement direction $\theta_m$.
An illustration of this setup is shown in Fig.~\ref{fig:anglesOfCell}.

\begin{figure}[ht]
 	\centering
 	\includegraphics[width=\linewidth]{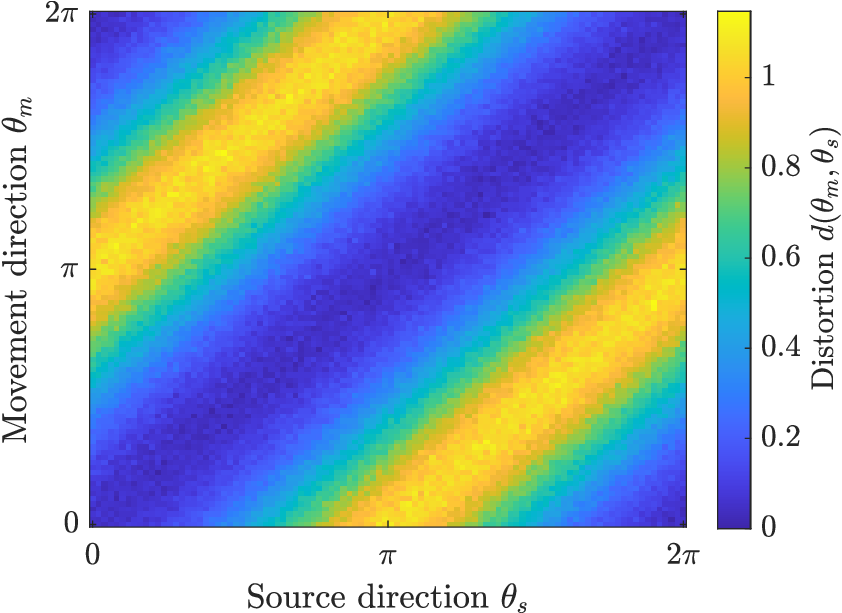}
 	\caption{\small Distortion function computed based on the \ac{IBAA} for \ac{LEGI} model with Hill coefficient $h=1$.}
 	\label{fig:distortionfunctionh1}
 \end{figure}

\begin{figure}[ht]
	\centering
	\includegraphics[width=\linewidth]{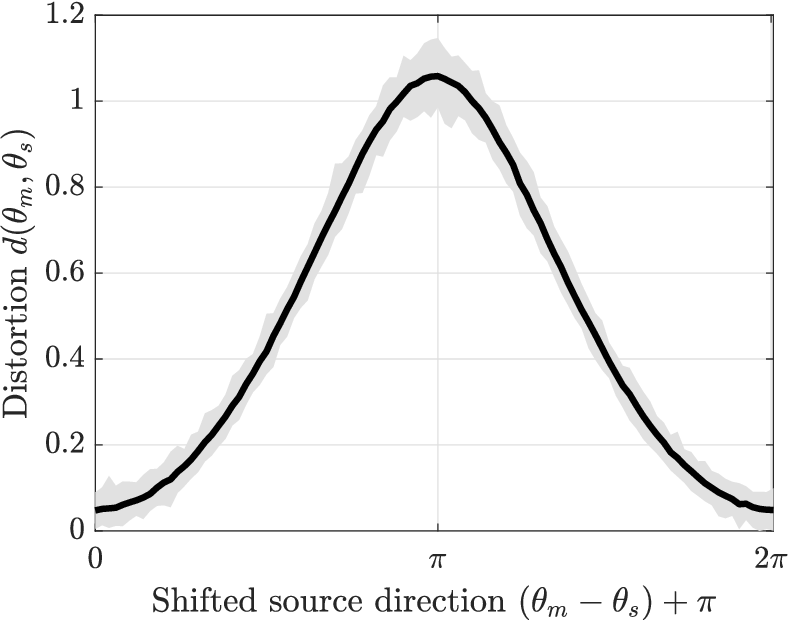}
	\caption{\small Distortion function computed from the simulation based on the \ac{LEGI} model with Hill coefficient $h=1$. The thick black line indicates the mean over all input values $\theta_s$.}
	\label{fig:shifteddistorion1}
\end{figure}

To generate the input, $\theta_\mathrm{s}$, output, $\theta_\mathrm{m}$, data of chemotaxis based on the \ac{LEGI} model we considered different values of Hill coefficient $h\in\{1,3,5,\ldots,15\}$, were higher values indicate higher amplification of the signal by the cell. For each fixed value of Hill coefficient we perform the simulation for $2\times10^7$ times. 
According to the \ac{IBAA} discussed in Section~\ref{sec:IBAA} we use Laplace smoothing~\eqref{eq:laplace} and Bayes' rule, and calculate the decision strategy $P_{\Theta_m|\Theta_s}(\theta_m|\theta_s)$, and compute distortion function $d(\theta_m,\theta_s)$.
 \begin{figure}[t]
 	\centering
 	\includegraphics[width=\linewidth]{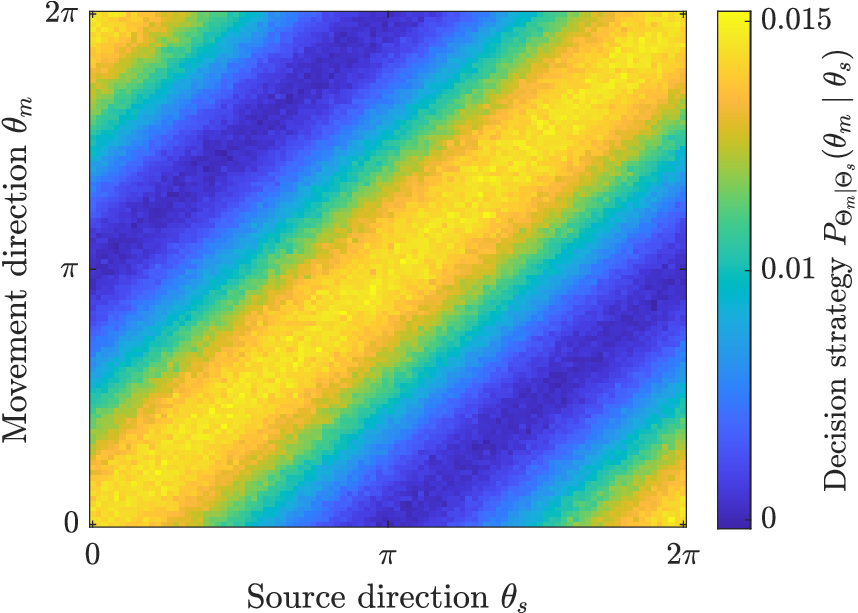}
 	\caption{\small Decision strategy $P_{\Theta_m|\Theta_s}(\theta_m|\theta_s)$ calculated from the \ac{LEGI} model for a Hill coefficient $h=1$.}
 	\label{fig:decisionstrategyh1}
 \end{figure}
Figure~\ref{fig:decisionstrategyh1} illustrates the decision strategy of the \ac{LEGI} model with Hill coefficient $h=1$. It shows the probability that the cell chooses to move in direction $\theta_m$ for a given source direction $\theta_s$.
The corresponding distortion function is shown in Fig.~\ref{fig:distortionfunctionh1}. We notice that if the decision angle $\theta_m$ is similar to the source direction $\theta_s$ the distortion is low. 
To facilitate comparison of the distortion function across different values of $\theta_s$, we apply a cyclic shift of $(\theta_m - \theta_s) + \pi$ to the function $d(\theta_s,\theta_m)$. This procedure aligns the functions such that it allows a direct comparison of their shapes.
The cyclic shifted distortion function is shown in Fig.~\ref{fig:shifteddistorion1} for $h=1$. In this figure the average distortion function looks similar to the theoretical cosine distortion function considered in~\cite{Andrews2007,Iglesias2016}.

\begin{figure}[ht]
    \centering
    \includegraphics[width=\linewidth]{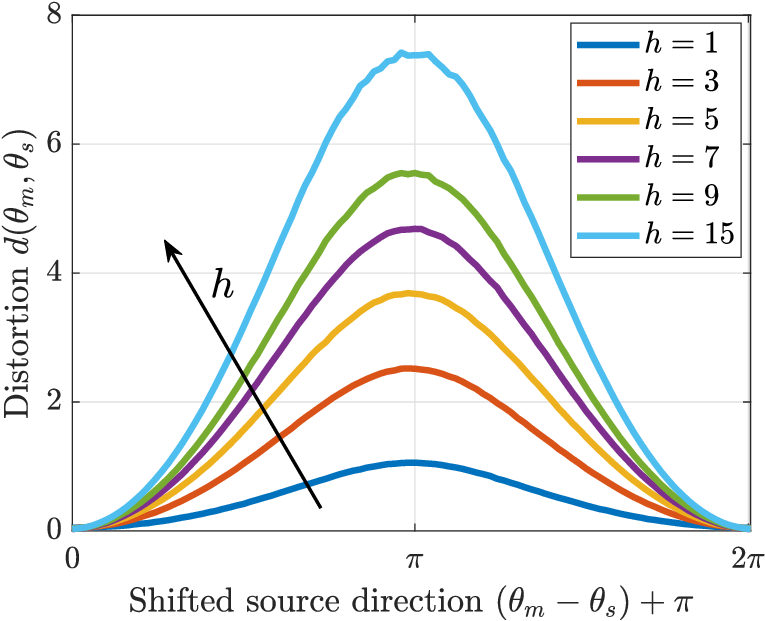}
    \caption{\small Distortion function corresponding to the \ac{LEGI} model for different values of Hill coefficient.}
    \label{fig:totalcosineabsolut}
\end{figure}

Figure~\ref{fig:totalcosineabsolut} shows the mean of the computed distortion functions for different Hill coefficients. Increasing the Hill coefficient leads to distortion functions of greater magnitude and sharper slope, indicating that the system assigns a higher penalty to potential errors. This trend is consistent with the findings in~\cite{Andrews2007}, which showed that a larger Hill coefficient has an effect analogous to increasing the parameter $\lambda$ in the \ac{BAA}, thereby enhancing the achievable mutual information. 

We observe that varying the Hill coefficient effectively alters the cell’s decision strategy. Biologically, the Hill coefficient reflects the degree of signal amplification in intracellular pathways, with higher values corresponding to stronger ultrasensitivity in the chemotactic response. This modulation can be interpreted as a change in the cell’s signaling amplification state, which manifests as a modification of the distortion function and results in higher mutual information. Prior work has typically assumed a fixed distortion function, independent of the cell’s internal state, and interpreted the Hill coefficient solely as a parameter governing the decision strategy. In contrast, we propose the concept of a \emph{state-dependent distortion function}, where the shape of the distortion function itself changes with the cell’s physiological condition. This perspective provides a foundation for modeling adaptive information-processing agents, in which distortion functions can be assumed to evolve in time.

In general, the \ac{IBAA} enables us to infer the underlying criteria that an agent uses to perform a given task in a noisy environment. Understanding these criteria (distortion function) provides insight into how natural systems make robust decisions, and may guide the design of artificial agents that replicate or draw inspiration from these biologically evolved criteria.

\section{Conclusion}\label{sec:conclusion}
In this paper, we model the chemotaxis process as an information processing system and propose the \ac{IBAA} to compute the distortion function from the cell's perspective. Using a cellular apoptosis model as a case study, we validate the \ac{IBAA} by computing distortion functions that accurately match their theoretical counterparts, based on input output data from simulation results. In the context of chemotaxis, we demonstrate a novel interpretation, distinct from prior work, showing that the cell's distortion function varies with the Hill coefficient, which is tied to the cell's state. This finding suggests that the cell's decision-making criteria adapt dynamically to the cell's state, enabling the exploration of time varying scenarios in future research. The proposed \ac{IBAA} offers a general framework for computing distortion functions across diverse systems, extending its applicability beyond chemotaxis to other biological and engineered systems requiring efficient information processing under uncertainty. Future work will focus on applying the \ac{IBAA} to dynamic and complex scenarios to further elucidate adaptive decision making processes for any types of agents.
%
%
\section*{Acknowledgment}
We would like to acknowledge Professor Pablo A. Iglesias for his helpful suggestions, which provided an essential foundation for this research.\\
This work is partly funded by the Deutsche Forschungsgemeinschaft (DFG, German Research Foundation) – GRK 2950 – Project-ID 509922606, and the European Union’s Horizon Europe – HORIZON-EIC-2024-PATHFINDEROPEN-01 under grant agreement Project N. 101185661.

\ifCLASSOPTIONcaptionsoff
  \newpage
\fi
\bibliographystyle{IEEEtran}
\bibliography{Bibliography}

\end{document}